\documentclass[12pt,a4paper]{article}
\usepackage{amsmath}
\usepackage{amssymb}

 \newcommand{\bs}[1]{\boldsymbol{#1}}

 \newcommand{\NN}{\mathbb N}
 \newcommand{\RR}{\mathbb R}
 \newcommand{\CC}{\mathbb C}
 \newcommand{\ZZ}{\mathbb Z}
 \newcommand{\II}{\mathbb I}
 
 \newcommand{\QQ}{\mathbb Q}
 \newcommand{\qed}{\hfill $\square$}

 \newtheorem{theorem}{Theorem}
 \newtheorem{lemma}{Lemma}
 \newtheorem{prop}{Proposition}

 \input{epsf}
 
\begin{document}
 \bibliographystyle{unsrt}

 \parindent0pt

\begin{center}
{\Large \bf Diffraction of random tilings:}

\bigskip
{\Large \bf some rigorous results}

 \end{center}
 \vspace{3mm}

 \begin{center}  {\sc Michael Baake} and {\sc Moritz H\"offe}
 
\vspace{5mm}

 Institut f\"ur Theoretische Physik, Universit\"at T\"ubingen, \\

 Auf der Morgenstelle 14, D-72076 T\"ubingen, Germany

\end{center}


\vspace{10mm}
\begin{abstract} 
The diffraction of stochastic point sets, both Bernoulli and Markov, and of 
random tilings with crystallographic symmetries is investigated in rigorous 
terms. In particular, we derive the diffraction spectrum of 1D random tilings, 
of stochastic product tilings built from cuboids, and of planar random tilings 
based on solvable dimer models, augmented by a brief outline of the diffraction 
from the classical 2D Ising lattice gas. We also give a summary 
of the measure theoretic approach to mathematical diffraction theory which 
underlies the unique decomposition of the diffraction spectrum into its pure 
point, singular continuous and absolutely continuous parts.
\end{abstract}

\vspace{5mm}

\parindent15pt
\vspace{5mm}

\subsection*{Introduction}

The diffraction theory of crystals is a subject with a long history, and one
can  safely say that it is well understood \cite{Guinier,Cowley}. Even though
the advent of quasicrystals, with their sharp diffraction images with perfect
non-crystallographic symmetry, seemed to question the general understanding,
the diffraction theory of perfect quasicrystals, in terms of the cut and project
method, is also rather well understood by now, see \cite{Hof,Hof-Waterloo} and
references therein. It should be noted though that this extension was by no
means automatic, and required a good deal of mathematics to clear up the
thicket. More recently, this has found a general extension to the setting of
locally compact Abelian groups \cite{Martin,Martin98} which can be seen as
a natural frame for mathematical diffraction theory and covers quite a number
of interesting new cases \cite{BMS}.

Another area with a wealth of knowledge is the diffraction theory of imperfect
crystals and amorphous bodies \cite{Guinier,Welberry}, but 
the state of affairs here is a lot less rigorous, and many results and
features seem to be more or less folklore. For example, the diffraction of
simple stochastic systems, as soon as they are not bound to a lattice, is 
only in its infancy, see \cite{BM1} for some recent addition to its rigorous
treatment. This does not mean that one would not know what to expect. However, 
one can often only find a qualitative argument in the literature, but no proof. 
May this be acceptable from a practical angle, it seems rather unsatisfactory 
from a more fundamental point of view. In other words, the answer to the 
question which distributions of matter diffract is a lot less known than one 
would like to believe, compare the discussion in \cite[Sec.\ 6]{Hof}
and also \cite{R2,EM}. 

Note that this question contains several different aspects. On the one hand,
one would like to know, in rigorous terms, under which circumstances the 
diffraction image is well defined in the sense that it has a unique infinite
volume limit. 
This is certainly the case if one can refer to the ergodicity of the underlying 
distribution of scatterers \cite{Dworkin,Hof,Martin98}, in particular, if their 
positional arrangement is linearly repetitive \cite{LP}.
However, this is often difficult to assess in situations {\em without}
underlying ergodicity properties, see \cite{BMP} for an example.
On the other hand, even if the image is uniquely defined, one still wants
to know whether it contains Bragg peaks or not, or if there is any 
diffuse scattering present in it.

This situation certainly did not improve with the more detailed investigation 
of quasicrystals, e.g.\ their less perfect versions, and in particular with the
study of the so-called random tilings \cite{Elser,Henley,Richard}. Again,
there is a good deal of folklore available, and a careful reasoning based
upon scaling arguments (compare \cite{Jaric,Henley}) seems to give convincing 
and rather consistent results on their diffraction properties.
However, various details, and in particular the exact nature of the diffraction
spectrum, have always been the topic of ongoing discussion, so that a more
rigorous treatment is desirable. It is the aim of this contribution to go one
step into this direction, and to extend the analysis of \cite{BM1} on 
generalized lattice gases to the case of certain Markov type systems as they 
appear in the theory of random tilings. 
We will not be able to answer the real questions
concerning those tilings relevant for quasicrystals, but we still think that
the results derived below are a worth-while first step.
Even this requires a bunch of methods and results which are
scattered over rather different branches of mathematics and mathematical 
physics. It is thus also one of our aims to recollect the essential aspects 
and references, tailored for what we need here and for future work in this 
direction.

Let us summarize how the article is organized. We start with a recapitulation 
of the measure theoretic setup needed for mathematical diffraction theory, where
we essentially follow Hof \cite{Hof,Hof-Waterloo}, but adapt and extend it to 
our needs. We will be a little bit more explicit here than needed for an 
audience with background in mathematics or mathematical physics, because
we hope that the article becomes more self-contained that way, and hence more
readable for physicists and crystallographers who usually do not approach 
problems of diffraction theory in these more rigorous terms. We consider this 
as part of an attempt to penetrate the communication barrier.
We then investigate several 1D systems, notably Markov systems and 1D random 
tilings, and derive their spectral properties. 
This is followed by an intermediate discussion of stochastic product tilings in 
arbitrary dimensions which already indicates that the appearance of mixed 
spectra with pure point, singular continuous and absolutely continuous parts 
is generic, though it also shows that its meaning in more than one dimension 
will have many facets. 

The next Section then deals with
the main results of this article, the derivation of the diffraction spectrum
of certain crystallographic random tilings in the plane, namely the domino and 
the rhombus (or lozenge) tiling. 
This requires some adaptation of results from the theory of Gibbs states
to special hard-core lattice systems. Again, we explain that in slightly
more detail than necessary from the point of view of mathematical physics
in order to enhance self-containedness and readability.
In both cases, the basic input of the explicit result has long been known
in statistical mechanics, but the interest in the diffraction issue is rather 
recent. We also briefly comment on the diffraction of an interactive lattice 
gas based upon the classical 2D Ising model and its implications. 
The discussion addresses some
open questions and what one should try to achieve next.

\subsection*{Recollections from mathematical diffraction theory}

Diffraction problems have many facets, but one important question certainly
is which distributions of atoms lead to well-defined diffraction images,
and if so, to what kind of images. This is a difficult problem, far from
being solved. So, one often starts, as we will also do here, by looking
at ``diffraction at infinity'' from single-scattering where it essentially 
reduces to questions
of Fourier analysis \cite[Sec.\ 6]{Arsac}. This is also called kinematic
diffraction in the Fraunhofer picture \cite{Cowley}, and we are looking into
the more mathematical aspects of that now.
Mathematical diffraction theory, in turn, is concerned with spectral 
properties of the Fourier transform of the autocorrelation measure of unbounded
complex measures. Let us therefore first introduce and discuss the notions 
involved. Here, we start from the presentation in \cite{Hof,Hof-Waterloo} 
where the linear functional approach to measures is taken, compare \cite{Dieu} 
for details and background material. We also introduce our notation this way.

Let ${\cal K}$ be the space of complex-valued continuous functions with compact 
support. A (complex) {\em measure} $\mu$ on $\RR^n$ is a linear functional on 
${\cal K}$ with the extra condition that for every compact subset $K$ of $\RR^n$ 
there is a constant $a^{}_K$ such that
\begin{equation}
     |\mu(f)| \; \leq \; a^{}_K \, \|f\|
\end{equation}
for all $f\in{\cal K}$ with support in $K$; here, $\|f\| = \sup_{x\in K} |f(x)|$
is the supremum norm of $f$. If $\mu$ is a measure, the {\em conjugate} of $\mu$
is defined by the mapping $f\to\overline{\mu(\bar{f})}$. It is again  
a measure and denoted by $\bar{\mu}$. A measure $\mu$ is called {\em real} (or
signed), if $\bar{\mu}=\mu$, or, equivalently, if $\mu(f)$ is real for all
real-valued $f\in{\cal K}$.
A measure $\mu$ is called {\em positive} if $\mu(f)\geq 0$ for all $f\geq 0$.
For every measure $\mu$, there is a smallest positive measure, denoted by 
$|\mu|$, such that $|\mu(f)|\leq |\mu|(f)$ for all non-negative $f\in{\cal K}$,
and this is called the {\em absolute value} (or the total variation) of $\mu$.

A measure $\mu$ is {\em bounded} if $|\mu|(\RR^n)$ is finite (with obvious
meaning, see below), otherwise it is called unbounded. Note that a measure 
$\mu$ is continuous on ${\cal K}$ with respect to the topology induced by the 
norm $\|.\|$ if and only if it is bounded \cite[Ch.\ XIII.20]{Dieu}. In view of
this, the vector space of measures on $\RR^n$, ${\cal M}(\RR^n)$, is given the 
{\em vague topology}, i.e.\ a sequence of measures $\{\mu_n\}$ converges
vaguely to $\mu$ if $\lim_{n\to\infty} \mu_n(f) = \mu(f)$ in $\CC$ for all
$f\in{\cal K}$. This is just the weak-* topology on ${\cal M}(\RR^n)$, 
in which all the ``standard'' linear operations on measures are continuous, 
compare \cite[p.\ 114]{RS} for some consequences of this. 
The measures defined this way are, by  proper decomposition
\cite[Ch.\ XIII.2 and Ch.\ XIII.3]{Dieu} and an application of the Riesz-Markov 
representation theorem, see \cite[Thm.\ IV.18]{RS} or 
\cite[Thm.\ 69.1]{Berberian}, in one-to-one correspondence with the regular 
Borel measures on $\RR^n$, wherefore we identify them. In particular, we write
$\mu(A)$ (measure of a set) and $\mu(f)$ (measure of a function) for simplicity.

{}For any function $f$, define $\tilde{f}$ by $\tilde{f}(x):=\overline{f(-x)}$.
This is properly extended to measures via 
$\tilde{\mu}(f):=\overline{\mu(\tilde{f})}$.
Recall that the convolution $\mu * \nu$ of two measures $\mu$ and $\nu$ is
given by $\mu * \nu (f) := \int f(x+y) \mu(dx) \nu(dy) $ which is well-defined
if at least one of the two measures has compact support. For $R>0$, let $B^{}_R$
denote the closed ball of radius $R$ with centre 0, and ${\rm vol}(B^{}_R)$ 
its volume. 
The characteristic function of a subset $A\subset\RR^n$ is denoted by $1^{}_A$.
Let $\mu^{}_R$ be the restriction of a measure $\mu$ to the ball $B^{}_R$.
Since $\mu^{}_R$ then has compact support,
\begin{equation}
        \gamma^R \; := \; \frac{1}{{\rm vol}(B^{}_R)} \, 
                          \mu^{}_R * \tilde{\mu}^{}_R
\end{equation}
is well defined. Every vague point of accumulation of $\gamma^R$, as 
$R\to\infty$, is called an {\em autocorrelation} of $\mu$, and as such it is, 
by definition, a measure. If only one point of accumulation exists, the 
autocorrelation is unique, and
it is called the {\em natural autocorrelation}. It will be denoted by $\gamma$
or by $\gamma^{}_{\mu}$ to stress the dependence on $\mu$. 
One way to establish the existence of the limit is through the pointwise
ergodic theorem, compare \cite{Dworkin}, if such methods apply. If not,
explicit convergence proofs will be needed, as is apparent from known
examples \cite{BMP} and counterexamples \cite{LP}.

Note that Hof \cite{Hof} uses cubes rather than balls in his definition of
$\gamma^R$. This simplifies some of his proofs technically, but 
they also work for balls which are more natural objects in a physical context.
This is actually not important for our purposes here. One should keep in mind,
however, that the autocorrelation will, in general, depend on the shape of the 
volume over which the average is taken --- with obvious meaning for the
experimental situation where the shape corresponds to the aperture. 
To get rid of this problem, one often restricts the class of models
to be considered and defines the limits over van Hove patches, 
thus demanding a stricter version of uniqueness \cite[Sec.\ 2.1]{Ruelle}.

The space of complex measures is much too general for our aims, and we have to
restrict ourselves to a natural class of objects now. A measure $\mu$ is called
{\em translation bounded} \cite{AG} if for every compact set $K\subset\RR^n$
there is a constant $b^{}_K$ such that
\begin{equation}
   \sup_{x\in\RR^n} |\mu|(K+x) \; \leq \; b^{}_K \, .
\end{equation}
For example, if $\Lambda$ is a point set of {\em finite local complexity}, 
i.e.\ if the set $\Delta=\Lambda-\Lambda$ of difference vectors is discrete 
and closed, the weighted Dirac comb 
\begin{equation} \label{comb1}
    \omega^{}_{\Lambda} \; := \; \sum_{x\in\Lambda} w(x) \delta_x \, ,
\end{equation}
where $\delta_x$ is Dirac's measure at point $x$, is certainly translation 
bounded if the $w(x)$ are complex numbers with 
$\sup_{x\in\Lambda} |w(x)| < \infty$.
This is so because $\Delta$ discrete and closed implies that $0\in\Delta$ is
isolated and the points of $\Lambda$ are separated by a minimal distance,
hence $\Lambda$ is uniformly discrete.
Translation bounded measures $\mu$ have the property that all $\gamma^R$ are
uniformly translation bounded, and if the natural autocorrelation exists,
it is clearly also translation bounded \cite[Prop.\ 2.2]{Hof}. This is a very
important property, upon which a fair bit of our later analysis rests.
Note that such a restriction is neither necessary, nor even desirable (it
would exclude the treatment of gases and liquids), but it is fulfilled in
all our examples and puts us into a good setting in all cases where we
cannot directly refer to pointwise ergodic theorems. Let us finally 
mention that different measures can lead to the same natural autocorrelation,
namely if one adds to a given measure $\mu$ a sufficiently ``meager'' measure
$\nu$, see \cite[Prop.\ 2.3]{Hof} for details. In particular, adding or removing
finitely many points from $\Lambda$, or points of density 0, does not change
$\gamma$, if it exists.

Let us focus on the Dirac comb $\omega = \omega^{}_{\Lambda}$ 
from (\ref{comb1}), with $\Lambda$ of finite local complexity, and let us 
assume for the moment that its natural autocorrelation $\gamma^{}_{\omega}$ 
exists and is unique (Lagarias and Pleasants construct an example where this 
is not the case \cite{LP}). A short calculation shows that
$\tilde{\omega}^{}_{\Lambda} = \sum_{x\in\Lambda} \overline{w(x)} \delta_{-x}$.
Since $\delta_x * \delta_y = \delta_{x+y}$, we get
\begin{equation} \label{comb2}
   \gamma^{}_{\omega} \; = \; \sum_{z\in\Delta} \nu(z) \delta_z \, ,
\end{equation}
where the autocorrelation coefficient $\nu(z)$, for $z\in\Delta$, is given 
by the limit
\begin{equation}
   \nu(z) \; = \; \lim_{R\to\infty} \frac{1}{{\rm vol}(B_R)}
          \sum_{\stackrel{\scriptstyle y \in \Lambda_R}
                         {\scriptstyle z-y \in \Lambda}}
               w(y) \, \overline{w(z-y)} \, ,
\end{equation}
where $\Lambda_R = \Lambda \cap B_R$. Conversely, if these limits exist for
all $z\in\Delta$, the natural autocorrelation exists, too, because $\Delta$ is
discrete and closed by assumption, and (\ref{comb2}) thus uniquely defines a
translation bounded measure of positive type. This is one advantage of using
sets of finite local complexity.

We now have to turn our attention to the Fourier transform of unbounded measures
on $\RR^n$ which ties the previous together with the theory of tempered
distributions \cite{Schwartz}, see \cite{AG,Martin98} for extensions to other 
locally compact Abelian groups.

Let ${\cal S}(\RR^n)$ be the space of rapidly decreasing functions
\cite[Ch.\ VII.3]{Schwartz}, also called Schwartz functions.
By the Fourier transform of a Schwartz function $\phi\in{\cal S}(\RR^n)$ we mean
\begin{equation} \label{fourierdef}
      ({\cal F}\phi) (k) \; = \;
      \hat{\phi} (k) \; := \; \int_{\RR^n} e^{-2\pi i k\cdot x} \, \phi(x) dx
\end{equation}
which is again a Schwartz function \cite{Schwartz,RS}. Here, $k\cdot x$ is
the Euclidean inner product of $\RR^n$. The inverse operation is given by
\begin{equation} \label{invfourierdef}
      \check{\psi} (x) \; = \; 
      \int_{\RR^n} e^{2\pi i x\cdot k} \, \psi(k) dk \, .
\end{equation}
The Fourier transform $\cal F$ is thus a linear bijection from 
${\cal S}(\RR^n)$ onto itself, and is bicontinuous \cite[Thm.\ IX.1]{RS}. 
Our definition (with the factor $2\pi$ in the exponent) results in the usual
properties, such as $\check{\hat{\phi}\,}=\phi$ and 
$\hat{\check{\psi}\,}=\psi$. 
The convolution theorem takes the simple form
$\widehat{\phi_1 * \phi_2} = \hat{\phi}_1 \cdot \hat{\phi}_2$ where convolution
is defined by
\begin{equation}
    \phi_1 * \phi_2 \, (x) \; := \; \int_{\RR^n} \phi_1(x-y)\phi_2(y)dy \, .
\end{equation}
Let us also mention that $\cal F$ has a unique
extension to the Hilbert space $L^2(\RR^n)$, often called the Fourier-Plancherel
transform, which turns out to be a unitary operator of fourth order, i.e.\
${\cal F}^4=\mbox{Id}$. This is so because $({\cal F}^2\phi)(x)=\phi(-x)$, see
\cite{Rudin} for details.

{}Finally, the matching definition of the Fourier transform of a tempered
distribution \cite{Schwartz} $T\in{\cal S}'(\RR^n)$ is
\begin{equation}
  \hat{T}(\phi ) \; := \; T( \hat{\phi} )
\end{equation}
for all Schwartz functions $\phi$, as usual. The Fourier transform is then a 
linear bijection of ${\cal S}'(\RR^n)$ onto itself which is the unique weakly 
continuous extension of the Fourier transform on ${\cal S}(\RR^n)$ 
\cite[Thm.\ IX.2]{RS}. This is important, because it means that weak 
convergence of a sequence of tempered distributions, $T_n\to T$ as 
$n\to\infty$, implies weak convergence of their Fourier transforms, 
i.e.\ $\hat{T}_n\to \hat{T}$.

Let us give three examples here, which will reappear later. 
First, the Fourier transform of Dirac's measure at $x$ is given by
\begin{equation}
   \hat{\delta}_x \; = \; e^{-2\pi i k\cdot x} 
\end{equation}
where the right hand side is actually the Radon-Nikodym density, and hence
a function of the variable $k$, that represents
the corresponding measure (we will not distinguish a measure from its
density, if misunderstandings are unlikely).
Second, consider the Dirac comb $\omega^{}_{\Gamma}=\sum_{x\in\Gamma}\delta_x$
of a lattice $\Gamma\subset\RR^n$ (i.e.\ a discrete subgroup of $\RR^n$ such
that the factor group $\RR^n/\Gamma$ is compact). Then, one has
\begin{equation}
  \hat{\omega}^{}_{\Gamma} \; = \; {\rm dens}(\Gamma) \cdot 
                                    \omega^{}_{\Gamma^*} \, ,
\end{equation}
where ${\rm dens}(\Gamma)$ is the density of $\Gamma$, i.e.\ the number of 
lattice points per unit volume, and $\Gamma^*$ is the {\em dual} 
(or reciprocal) lattice,
\begin{equation} \label{psf}
    \Gamma^* \; := \; \{ y\in\RR^n\mid x\cdot y \in\ZZ \mbox{ for all }
                         x\in\Gamma \} \, .
\end{equation}
This is {\em Poisson's summation formula} for distributions 
\cite[p.\ 254]{Schwartz}
and will be central for the determination of the Bragg part of the diffraction
spectrum. Finally, putting these two pieces together, we also get the formula
\begin{equation} \label{exp-sum}
       \sum_{x\in\Gamma}  e^{-2\pi i k\cdot x} \; = \;
       {\rm dens}(\Gamma) \cdot \sum_{y\in\Gamma^*} \delta_y \, ,
\end{equation}
to be understood in the distribution sense.

If a measure $\mu$ defines a tempered distribution $T_{\mu}$ by
$T_{\mu}(\phi) = \mu(\phi)$ for all $\phi\in{\cal S}(\RR^n)$, the measure is
called a {\em tempered measure}. A sufficient condition for a measure to be
tempered is that it increases only slowly, in the sense that
$\int (1+|x|)^{-\ell} |\mu|(dx) < \infty$ for some $\ell\in\NN$, see
\cite[Thm.\ VII.VII]{Schwartz}. Consequently, every translation bounded
measure is tempered -- and such measures form the right class for our
purposes. We will usually not distinguish between a measure and the
corresponding distribution, i.e.\ we will write $\hat{\mu}$ for $\hat{T}_{\mu}$.
The Fourier transform of a tempered measure is a tempered distribution, but it
need {\em not} be a measure. However, if $\mu$ is of {\em positive type} (also
called positive definite) in the sense that $\mu(\phi*\tilde{\phi})\geq 0$
for all $\phi\in{\cal S}(\RR^n)$, then $\hat{\mu}$ is a positive measure
by the Bochner-Schwartz Theorem \cite[Thm.\ IX.10]{RS}. Every autocorrelation
$\gamma$ is, by construction, a measure of positive type, so that $\hat{\gamma}$
is a positive measure. This explains why this is a natural approach to kinematic
diffraction, because the observed intensity pattern is represented by a positive
measure that tells us which amount of intensity is present in a given volume.

Also, taking Lebesgue's measure as a reference, positive measures $\mu$ admit a 
unique decomposition into three parts,
\begin{equation}
 \mu \; = \; \mu^{}_{pp} + \mu^{}_{sc} + \mu^{}_{ac} \, ,
\end{equation}
where $pp$, $sc$ and $ac$ stand for pure point, singular continuous and
absolutely continuous, see \cite[Sec.\ I.4]{RS} for background material.
The set $P=\{x\mid\mu(\{x\})\neq 0\}$ is called the set of pure points of $\mu$,
which supports the so-called Bragg part $\mu^{}_{pp}$ of $\mu$.
Note that $P$ is at most a countable set. The rest, i.e.\ $\mu - \mu^{}_{pp}$,
is the ``continuous background'' of $\mu$, and this is the unambiguous and 
mathematically precise formulation of what such terms are supposed to mean.
Depending on the context, one also writes
\begin{equation}
   \mu \; = \; \mu^{}_{pp} + \mu^{}_{cont}
       \; = \; \mu^{}_{sing} + \mu^{}_{ac} \, ,
\end{equation}
where $\mu^{}_{cont}=\mu^{}_{sc}+\mu^{}_{ac}=\mu-\mu^{}_{pp}$ is the continuous
part of $\mu$ (see above) and $\mu^{}_{sing}=\mu^{}_{pp}+\mu^{}_{sc}$ is the 
singular part, i.e.\ $\mu^{}_{sing}(S)=0$ for some set $S$ whose complement has 
vanishing Lebesgue measure (in other words, $\mu^{}_{sing}$ is concentrated to 
a set of vanishing 
Lebesgue measure). Finally, the absolutely continuous part, which is called
diffuse scattering \cite{JF} in crystallography, can be represented by 
its Radon-Nikodym density \cite[Thm.\ I.19]{RS} which is often very handy. 
Examples for the various spectral types can easily be constructed by different 
substitution systems, see \cite{Queffelec} and references therein for details.
Later on, we shall meet a simple example in the context of stochastic product
tilings where all three spectral types are present, though their meaning will 
need a careful discussion.

Hof discusses a number of properties of Fourier transforms of tempered measures
\cite{Hof,Hof-Waterloo}. Important for us is the observation that temperedness
of $\mu$ together with positivity of $\hat{\mu}$ implies translation boundedness
of $\hat{\mu}$ \cite[Prop.\ 3.3]{Hof}. So, if $\mu$ is a translation bounded
measure whose natural autocorrelation $\gamma^{}_{\mu}$ exists, then
$\gamma^{}_{\mu}$ is also translation bounded (see above), hence tempered,
and thus the positive measure $\hat{\gamma}^{}_{\mu}$ is also both translation
bounded and tempered. 
This is the situation we shall meet throughout the article.

In what follows, we shall restrict ourselves to the spectral analysis of
measures $\mu$ that are concentrated on uniformly discrete point sets.
They are seen as an idealization of pointlike scatterers at uniformly
discrete positions, in the infinite volume limit.
The rationale behind this is as follows. If one understands these cases well,
one can always extend both to measures with extended local profiles (e.g.\ by
convolution of $\omega$ with a smooth function of compact support or with a
Schwartz function) and to measures that describe diffraction at positive
temperatures (e.g.\ by using Hof's probabilistic treatment \cite{Hof95}).
The treatment of gases or liquids might need some additional tools, but
we focus on situation that stem from solids with long-range order and
different types of disorder, because we feel that this is where the biggest
gaps in our understanding are at present.

\subsection*{Illustrative results in one dimension}

The simplest cases to be understood are those in one dimension. 
We start with some examples obtained from stationary stochastic processes and
then derive in detail the diffraction properties of 1D random tilings.
The language and methods of this section closely follow those of classical
ergodic theory because very good literature is available here \cite{Petersen}.

\subsubsection*{Bernoulli and Markov systems}

Let us start with a Bernoulli system, i.e.\ with a lattice gas without 
interaction.
\begin{prop} \label{bernoulli}
  Consider the stochastic Dirac comb $\omega = \sum_{m\in\ZZ} \eta(m) \delta_m$
  where $\eta(m)$ is a family of i.i.d.\ random variables that can take
  any of the $n$ complex numbers $h^{}_1, \ldots, h^{}_n$ (assumed pairwise
  different), with attached probabilities $p^{}_1, \ldots, p^{}_n$, $p^{}_i>0$.

  Then, the autocorrelation $\gamma^{}_{\omega}$ of $\omega$ exists with
  probabilistic certainty\footnote{Here and in the sequel, assertions of
  probabilistic certainty always refer to the invariant measure
  of the corresponding stochastic process.} and has the form 
  $\gamma^{}_{\omega} = \sum_{m\in\ZZ} \nu(m) \delta_m$ with autocorrelation
  coefficients
\begin{equation}
    \nu(m) \; = \; \begin{cases} 
                      \langle |\bs{h}|^2 \rangle \, , & \text{if $m=0$}, \\
                     |\langle \bs{h} \rangle|^2 \, , & \text{if $m\neq 0$}, 
                   \end{cases}
\end{equation}
  where $\langle |\bs{h}|^2 \rangle = \sum_{i=1}^n p^{}_i |h^{}_i|^2$
  and $\langle \bs{h} \rangle =\sum_{i=1}^n h^{}_i p^{}_i$.
  Consequently, the diffraction measure is, with probability one, 
  $\ZZ$-periodic and given by
\begin{equation}
   \hat{\gamma}^{}_{\omega} \; = \; 
        \left(|\langle \bs{h} \rangle|^2 \sum_{m\in\ZZ} \delta_m \right) 
        \; + \; 
        \left(\,\langle |\bs{h}|^2 \rangle - |\langle \bs{h} \rangle|^2 
        \,\right) \, .
\end{equation}
\end{prop}
{\sc Proof}: This is a straight-forward application of Birkhoff's pointwise 
ergodic theorem \cite[Thm.\ 2.3]{Petersen}, applied to the case of a Bernoulli 
system as described in \cite[Sec.\ 1.2 C]{Petersen}. One identifies the 
possible Dirac combs with the corresponding bi-infinite sequences 
$\bs{x}=(x^{}_i)^{}_{i\in\ZZ}$
of the Bernoulli process given above. Then, $\nu(m)$ is the orbit average
of the function $f(\bs{x}) = \bar{x}^{}_0 x^{}_m$ under the action of the shift
which almost surely exists and equals the average of $f$ over the invariant 
measure, because the process is ergodic. \qed \smallskip

The diffraction thus consists of a pure point part (Bragg peaks) that is the 
one of the regular lattice $\ZZ$ multiplied by the absolute square of the 
average scattering strength and an absolutely continuous part 
(diffuse scattering) which is 
constant in this case (hence it is ``white noise''), as one would expect for a 
Bernoulli process. Note that the entropy density of this ensemble is given by
$s=-\sum_{i=1}^n p_i\log(p_i)$, see \cite[Ch.\ 5.3, Ex.\ 3.4]{Petersen}.
One could, alternatively, refer to the strong law of large numbers, under 
slightly
different assumptions. This would then also give a generalization to higher
dimensions, and to regular point sets beyond lattices, see \cite{BM1} for
a detailed account of this.

Let us now turn our attention to stochastic systems with interaction. Let $M$ 
be a Markov (or stochastic) matrix, i.e.\ $M=(M_{ij})^{}_{1\leq i,j\leq n}$ 
with $M_{ij}\geq 0$ and $\sum_{j=1}^{n}M_{ij}=1$. We assume that $M$ 
is {\em primitive},
so some power of $M$ has strictly positive entries only. As a consequence,
the Perron-Frobenius (PF) eigenvalue $\lambda_1=1$ is unique, and all other 
eigenvalues $\lambda_i$ 
of $M$ have absolute value $|\lambda_i|<1$. The corresponding right eigenvector
is $(1,1,\ldots,1)^t$, while the left eigenvector $\bs{p} = \bs{p} M$,
$\bs{p}=(p^{}_1,\ldots,p^{}_n)$, defines the stationary state. Due to the
primitivity of $M$, we can choose $p^{}_i>0$ and statistical normalization, 
$\sum_{i=1}^{n} p^{}_i=1$. Let $\Pi=\mbox{diag}(p^{}_1,\ldots,p^{}_n)$
which is thus invertible.

In view of the applications we have in mind, we want to consider a Markov 
process that gives the same result for an operation in reverse direction. 
That is to say we restrict ourselves to {\em reversible} processes, i.e.\ 
to stochastic matrices $M$ with
\begin{equation} \label{reversible}
        \Pi M \; = \; M^t \Pi \, .
\end{equation}
Since $M$ is also primitive, all $p^{}_i>0$ and $P=\Pi^{1/2}$ is well defined
and non-singular, $P=\mbox{diag}(\sqrt{p^{}_1},\ldots,\sqrt{p^{}_n}\,)$.
Now, $S=PMP^{-1}$ is a real symmetric matrix and can thus be diagonalized by
an orthogonal matrix. The eigenvalues of $S$ and $M$ coincide and are real; 
we denote them by $\lambda_1=1, \lambda_2,\ldots,\lambda_n$, where 
$|\lambda_i|<1$ for all $i\geq 2$. If $\{\bs{b}_i\}_{1\leq i\leq n}$ is the 
corresponding orthonormal basis, the spectral decomposition of $S$ is
\begin{equation}\label{decomp}
      S \; = \;  |\bs{b}_1\rangle \langle\bs{b}_1| \oplus
                 \sum_{i=2}^{n} |\bs{b}_i\rangle \lambda_i\langle\bs{b}_i|
        \; = \;  S_0 \oplus S_1 \, ,
\end{equation}
where we use Dirac's bra-ket notation for the standard Hermitian scalar product
of $\CC^n$. We embed $\RR^n$ into $\CC^n$ because we deal with complex 
scattering strengths later. Note that the decomposition on the right hand side 
of (\ref{decomp}) is into a projector (first term, $S_0$) and a contraction,
i.e.\  $|S_1 \bs{x}| < |\bs{x}|$ for all $\bs{x}\in\CC^n$. Also, we have
$S_0 S_1 = S_1 S_0=0$ and, in the standard basis of $\CC^n$, $S_0$ is explicitly
given by $S_0=(\sqrt{p^{}_i p^{}_j}\,)^{}_{1\leq i,j \leq n}$.

\begin{prop} \label{markov}
 Consider the stochastic Dirac comb $\omega = \sum_{m\in\ZZ} \eta(m) \delta_m$
 where $\eta(m)$ is a family of random variables that take values out of
 the $n$ complex numbers $h^{}_1, \ldots, h^{}_n$ (assumed pairwise
 different), subject to a primitive, reversible Markov process defined
 by a matrix $M$ with left-PF-eigenvector $\bs{p}$ as described above.

 Then, the autocorrelation $\gamma^{}_{\omega}$ of $\omega$ exists with
 probabilistic certainty and has the form 
 $\gamma^{}_{\omega} = \sum_{m\in\ZZ} \nu(m) \delta_m$ with non-negative
 autocorrelation coefficients
\begin{equation}
    \nu(m) \; = \; \langle\bs{h}|\Pi M^{|m|}|\bs{h}\rangle
\end{equation}
 for $m\in\ZZ$. In particular, 
 $\nu(0)=\langle|\bs{h}|^2\rangle=\sum_{i=1}^{n} p^{}_i |h^{}_i|^2$.
\end{prop}
{\sc Proof}: This is another application of Birkhoff's pointwise ergodic
theorem \cite[Thm.\ 2.3]{Petersen}, this time applied to the case of an
ergodic Markov system as described in \cite[Sec.\ 1.2 D]{Petersen}. The
setup is parallel to that of the Bernoulli system, only the invariant measure
(defined via cylinder sets) is different, and this accounts for the different
ensemble average. The latter is calculated as (for $m\geq 0$ say)
$$ \nu(m) \; = \; 
   \sum_{i^{}_0,i^{}_1,\ldots,i^{}_m}
     \bar{h}^{}_{i^{}_0} p^{}_{i^{}_0} M^{}_{i^{}_0 i^{}_1} M^{}_{i^{}_1 i^{}_2}
     \cdot \ldots \cdot M^{}_{i^{}_{m-1} i^{}_m} h^{}_{i^{}_m} 
     \; = \;  \langle \bs{h} | \Pi M^m | \bs{h} \rangle \, . $$
Due to (\ref{reversible}), we also have
$$ \langle \bs{h} | \Pi M^m | \bs{h} \rangle
   \; = \; \langle \bs{h} | (M^t)^m \Pi | \bs{h} \rangle
   \; = \; \langle \bs{h} | (\Pi M^m)^{\dagger} | \bs{h} \rangle
   \; = \; \overline{\langle \bs{h} | \Pi M^m | \bs{h} \rangle} \, ,$$
which shows that $\overline{\nu(m)}=\nu(m)$, so $\nu(m)$ is real. The case $m<0$
is analogous. But we also have
\begin{equation}
   \nu(m) \; = \; \langle \, \bs{h} \mid 
          \frac{1}{2} (\Pi M^{|m|} + (M^t)^{|m|}\Pi)
                  \mid \bs{h} \, \rangle
          \; \geq \; 0 \, ,
\end{equation}
because it represents a quadratic form with a non-negative real symmetric 
(hence Hermitian) matrix. Finally, $m=0$ gives the result for $\nu(0)$.  
\qed \smallskip

{}From the last Proposition, it is possible to derive the diffraction.
Observe that, with $\bs{c} = P \bs{h}$ and $r > 0$, we have
\begin{equation}
    \nu(r) \; = \; \langle \bs{c} | S^r | \bs{c} \rangle
           \; = \; \langle \bs{c} | S_0^r | \bs{c} \rangle +
                   \langle \bs{c} | S_1^r | \bs{c} \rangle \, ,
\end{equation}
and that $S_0$ is a projector, i.e.\ $S_0^r=S_0^{}$. With the explicit
form of $S_0$ given above, we thus obtain ($r > 0$)
\begin{equation}
    \nu(r) \; = \; \left|\sum_{i=1}^{n} p^{}_i h^{}_i\right|^2 +
                   \langle \bs{c} | S_1^r | \bs{c} \rangle 
           \; = \; |\langle\bs{h}\rangle|^2 +
                   \langle \bs{c} | S_1^r | \bs{c} \rangle \, .
\end{equation}
With $\nu(0) = \langle|\bs{h}|^2\rangle =
|\langle\bs{h}\rangle|^2 + (\langle|\bs{h}|^2\rangle-|\langle\bs{h}\rangle|^2)$,
the autocorrelation gets the form
\begin{equation}\label{markov-auto}
   \gamma^{}_{\omega} \; = \; |\langle\bs{h}\rangle|^2 \sum_{m\in\ZZ}\delta_m
       + (\langle|\bs{h}|^2\rangle-|\langle\bs{h}\rangle|^2)\delta_0
       + \sum_{r=1}^{\infty} \langle\bs{c}|S_1^r|\bs{c}\rangle
         (\delta_r + \delta_{-r}) \, .
\end{equation}

Before we continue, let us point out that the Bernoulli case is a special
case of this, namely $M_{ij}=1/n$, hence $p^{}_1=\ldots=p^{}_n=1/n$,
$S_0=M$ and $S_1=0$. In this limit, (\ref{markov-auto}) gives back the
corresponding result of Proposition \ref{bernoulli}. Also, one can treat
more general cases of Markov chains, compare \cite[vol.\ 1, Ch.\ XV]{Feller1} 
for details,
and Markov chains of higher order (or depth). This becomes technically 
more involved, but we think that the essential flavour is obvious from the
situation discussed here. It should also be clear how to make the other
examples of the crystallographic literature, see \cite{JF}, rigorous this way.
Finally, let us remark that the similarity of the (binary) 
Markov chain to the 1D Ising model is anything but accidental, and that 
the form of the autocorrelation is closely related to the solution of the Ising
model by means of transfer matrices, see \cite{Welberry} and
\cite[Ch.\ 3.2]{Georgii}.

The three terms on the right hand side of (\ref{markov-auto}) can now
easily be Fourier transformed. The first, by means of Poisson's summation
formula (\ref{psf}), gives the Bragg part. The second results in a constant 
continuous background, 
$\langle |\bs{h}|^2 \rangle - |\langle \bs{h} \rangle|^2$,
as in our previous example. To calculate the Fourier transform of the third 
term, we can employ Neumann's series \cite[p.\ 191]{RS} twice, because with 
$S_1$ also $\exp(\pm2\pi i k) S_1$ is a contraction, for arbitrary $k\in\RR$. 
This gives an absolutely continuous contribution which depends on the wave 
number $k$. Observing $\langle |\bs{h}|^2 \rangle = 
\langle \bs{h} | P^2 | \bs{h} \rangle$, one can combine the $k$-dependent
part with the first term of the constant part. This finally gives
\begin{theorem}
  The diffraction spectrum of the Markov system of Proposition \ref{markov}
  exists with probabilistic certainty and is given by the formula
\begin{equation}
  \hat{\gamma}^{}_{\omega} \; = \; |\langle\bs{h}\rangle|^2 \cdot 
      \omega^{}_{\ZZ} \; + \; (\hat{\gamma}^{}_{\omega})_{ac} \, ,
\end{equation}
  where $\omega^{}_{\ZZ}=\sum_{k\in\ZZ}\delta_k$ is the Dirac comb of the 
  integer lattice. The absolutely continuous part 
  $(\hat{\gamma}^{}_{\omega})_{ac}$ is
  represented by the $\ZZ$-periodic continuous (hence bounded) function
\begin{equation}\label{cont2}
  f(k) \; = \; \langle \bs{h} \mid P \,
       \frac{1 - S_1^2}{1 - 2 \cos(2\pi k) S_1^{} + S_1^2}
               \; P \mid \bs{h} \rangle \; - \; |\langle\bs{h}\rangle|^2 \, ,
\end{equation}
  with obvious meaning of the quotient.  \qed 
\end{theorem}

Let us add that an explicit calculation of $f(k)$ is easily done by means of
the orthonormal eigenbasis of $S$. If 
$P\bs{h}=\sum_{i=1}^{n} \beta_i \bs{b}_i$, 
one finds $|\beta_1|^2 = |\langle\bs{h}\rangle|^2$ and hence
\begin{equation} \label{basis-sum}
  f(k) \; = \;  \sum_{j=2}^{n} 
       \frac{|\beta_j|^2 \, (1 - \lambda_j^2)}
            {1 - 2 \cos(2\pi k) \lambda_j^{} + \lambda_j^2} 
       \; \geq \; 0 \, ,
\end{equation}
which, for $n=2$, coincides with the result of \cite{JF,Welberry}. 
Note that the diffuse background corresponds to a positive entropy density
which is given by $s=-\sum_{i,j} \, p_i M_{ij}\log(M_{ij})$, see
\cite[Ch.\ 5.3, Ex.\ 3.5]{Petersen}. Let us also mention that (\ref{cont2}) can
also be rewritten as
\begin{equation}
 f(k) \; = \; \langle \bs{h} \mid P \,
         \frac{1 - S_1^2 - S^{}_0}{1 - 2 \cos(2\pi k) S_1^{} + S_1^2}
         \; P \mid \bs{h} \rangle \, ,
\end{equation}
from which $f(k)\geq 0$ can easily be seen also in operator form.

{}For a graphical illustration of the diffraction, we refer to \cite{Welberry}.
The important observation is that, in the presence of an interaction, the
(now structured) diffuse background is attracted or repelled by the Bragg
peaks according to the interaction being attractive or repulsive. This is a
general qualitative feature of diffuse scattering, and will reappear in our
later examples.

\subsubsection*{1D random tilings}

Let us now change our point of view and consider a Bernoulli system not in
the scattering strength $h$ but rather in the distances. To keep things simple,
consider the case of placing two intervals, of length $u$ and $v$ (both $>0$), 
with probabilities $p$ and $q=1-p$ (hence, the entropy density per interval is 
$s=-p\log(p)-q\log(q)$), and assume that we have a unit point mass always at 
the left endpoint of each interval. Here, we have to distinguish the cases where
$\alpha = u/v$ is rational or irrational.

\begin{prop} \label{1drt}
  Consider the ensemble of binary random tilings of intervals of length $u$ and
  $v$, with probabilities $p$ and $q=1-p$, $pq>0$. Let, for each such random 
  tiling, $\Lambda$ be the point set defined through the left endpoints of the 
  intervals. Then, the natural density of $\Lambda$ exists with probabilistic 
  certainty and is given by $d=(pu+qv)^{-1}$.

  If $\omega=\omega^{}_{\Lambda}=\sum_{x\in\Lambda}\delta_x$ denotes the 
  corresponding stochastic Dirac comb, the autocorrelation 
  $\gamma^{}_{\omega}$ of $\omega$ also exists with probabilistic certainty. 
  It is a pure point measure that is supported on the set
\begin{equation} \label{diff1}
   \Delta \; = \; \{ m u + n v \mid m,n\in\ZZ \mbox{ and } mn\geq 0 \} \, ,
\end{equation}
  and, with $z_{m,n}:=m u + n v$, it is given by
\begin{equation}  \label{corr3}
   \gamma^{}_{\omega} \; = \; d \; \sum_{N=-\infty}^{\infty}
        \sum_{\ell=0}^{|N|} \binom{|N|}{\ell} p^{\ell} q^{|N|-\ell}
        \, \delta_{{\rm sgn}(N) z^{}_{\ell,|N|-\ell}} \, .
\end{equation}

  {}For $\alpha=u/v$ irrational, it is of the form
  $\gamma^{}_{\omega} = \sum_{z\in\Delta} \nu(z) \delta_z$, where
  $z\in\Delta$ has a unique representation $z=z_{m,n}=mu+nv$ with $m,n\in\ZZ$ 
  and $mn\geq 0$. The corresponding autocorrelation coefficient is then given by
\begin{equation} \label{coeff3}
  \nu(z_{m,n}) \; = \; d \, \binom{|n|\!+\!|m|}{|m|} \, p^{|m|} q^{|n|}  \, .
\end{equation}
\end{prop}
{\sc Proof}: Let $\Lambda$ be a random point set according to the assumptions.
If $u=v$, $\omega^{}_{\Lambda}$ is the Dirac comb of a lattice and 
the statement is trivial. So, let us henceforth assume that $u\neq v$.

If we view the random tiling ensemble as a Bernoulli system in the symbols 
$u,v$ with attached probabilities $p,q$, each tiling is a sequence 
$\bs{x}=(x^{}_i)^{}_{i\in\ZZ}$ with $x_i\in\{u,v\}$. 
Since $u,v$ also code the length of the intervals, the average distance between 
two consecutive points of the corresponding point set $\Lambda$ is the limit
of $\frac{1}{2n+1}\sum_{i=-n}^{n} x_i$ as $n\to\infty$ which almost surely
exists and, again by Birkhoff's pointwise ergodic theorem, is given by
$(pu+qv)$. The density $d$ is then clearly the inverse of this, as stated.

The possible differences between two points of $\Lambda$ are clearly given by
$\Delta$ of Eq.~(\ref{diff1}), and this set is discrete and closed. To 
establish the existence of the autocorrelation, it is thus sufficient to show 
that its coefficients exist. 
Let $z=mu+nv$ be in $\Delta$. Although the representation of $z$ need not
be unique (e.g.\ if $\alpha$ is rational), there is no other representation
with $N=m+n$ intervals because we have excluded the case $u=v$. So, starting
from an arbitrary point $x\in\Lambda$, $N+1$ different points can be reached by 
adding $N$ intervals (to the left or to the right, 
according to the sign of $z$), 
and the corresponding probabilities follow a binomial distribution, because the 
system is Bernoulli. If we pick a single sequence, and determine the average
frequency to reach the point $x+z$ from $x$ in $N$ steps, the strong law of
large numbers \cite[vol.\ 1, Ch.\ X.1]{Feller1} tells us that this frequency, 
almost surely, converges to the corresponding probability, i.e.\ we find the 
limiting frequency $$\binom{|n|+|m|}{|m|} p^{|m|} q^{|n|}\, .$$
Here, we have averaged over the number $M$ of starting points $x$ in a finite
piece of the sequence $\Lambda$, and considered the limit $M\to\infty$.
The corresponding contribution to the autocorrelation coefficient, however, 
is defined as a volume-averaged limit. This gives a prefactor that is the 
average number of points of $\Lambda$ per unit volume which is the density $d$. 
It exists with probability one as shown above. This establishes (\ref{corr3}).

The full autocorrelation coefficient is now given by  
\begin{equation}
   \nu(z) \; = \; \lim_{r\to\infty} \frac{1}{2r}
          \sum_{\stackrel{\scriptstyle x\in\Lambda, \, |x|\leq r}
                     {\scriptstyle x+z\in\Lambda}} 1
          \; = \; d \sum_{\stackrel{\scriptstyle m,n\in\ZZ, \, mn\geq 0}
                                     {\scriptstyle mu+nv=z}}
                      \binom{|n|+|m|}{|m|} p^{|m|} q^{|n|} \, ,
\end{equation}
which clearly also exists with probability one. This shows the existence of 
$\gamma^{}_{\omega}$. 

Let us finally assume $\alpha\not\in\QQ$.
Then, $z_{m,n}=z_{m',n'}$ implies $m=m'$ and $n=n'$, so that the only
possibility to fill this distance is by $m$ intervals of length $u$ and
$n$ of length $v$, the remaining freedom just being the order in which this is 
done. So, the sum in the previous equation reduces to one term, the one
given in (\ref{coeff3}).  \qed \smallskip

Let us add two remarks. First, the autocorrelation could also be worked 
out\footnote{We thank A.~Martin-L\"of for pointing this out to us.}
by means of the renewal theorem \cite[vol.\ 2, Ch.\ 11]{Feller1}. This would 
have the advantage
of also being applicable to gases. However, for our case, one has to pay
attention to convergence questions wherefore the derivation is not shorter.
Second, the argument given for the existence of the average distance
between two consecutive points of $\Lambda$ can easily be modified to calculate
that the average distance bridged by $N$ consecutive intervals is given by
$N(pu+qv)$, because the system is Bernoulli. This average, however, can now
also be calculated as the weighted sum over the possibilities to fill $N$ steps
by $m$ intervals of type $u$ and $N-m$ of type $v$, i.e.\ we obtain the identity
\begin{equation} \label{amaz1}
   \sum_{\ell=0}^{N} \binom{N}{\ell} p^{\ell} q^{N-\ell} \,
         (\ell u + (N-\ell) v) \; = \; N \, (pu+qv)
\end{equation}
which can also be checked explicitly by induction. It rests upon $p+q=1$, and 
the binomial formula for $(p+q)^N$.
It can also be understood from the first moment of the binomial distribution,
 calculated as derivative of its generating function.

To understand the diffraction, we have to determine the Fourier transform of 
$\gamma^{}_{\omega}$. To do so, it is advantageous to write the tempered 
measure $\gamma^{}_{\omega}$ as a weak limit of tempered measures with compact 
support, i.e.\ to write $\gamma^{}_{\omega} = \lim_{N\to\infty} \mu^{}_N$ where
\begin{equation}
    \mu^{}_N \; = \; d \; \sum_{n=-N}^{N}
        \sum_{m=0}^{|n|} \binom{|n|}{m} p^{m} q^{|n|-m}
        \, \delta_{{\rm sgn}(n) z^{}_{m,|n|-m}} \, ,
\end{equation}
with $z^{}_{m,n}=mu+nv$ as before. It is evident that this sequence of
measures converges weakly to $\gamma^{}_{\omega}$ of (\ref{corr3}), and
the support of $\mu^{}_N$ is certainly contained in the interval $[-w,w]$
where $w=N \max(u,v)$. So, due to the Paley-Wiener theorem 
\cite[Thm.\ IX.12]{RS},
the Fourier transform $\hat{\mu}^{}_N$ is naturally represented by an entire
analytic function, $g^{}_N(k)$. Also, since the Fourier transform is continuous,
the convergence of $\mu^{}_N\to\mu=\gamma^{}_{\omega}$ implies that of
$\hat{\mu}^{}_N\to\hat{\mu}=\hat{\gamma}^{}_{\omega}$. Note, however, that
the $\mu^{}_N$ are, in general, not measures of positive type, whence the 
$\hat{\mu}^{}_N$ are not positive measures. They are signed measures though, as 
we shall see shortly, and one could decompose them as 
$\hat{\mu}^{}_N=\hat{\mu}^{+}_N-\hat{\mu}^{-}_N$ with
$\hat{\mu}^{\pm}_N=\frac{1}{2}(|\hat{\mu}^{}_N|\pm \hat{\mu}^{}_N)$.
Then, $\hat{\mu}^{+}_N\to\hat{\gamma}^{}_{\omega}$ and
$\hat{\mu}^{-}_N\to 0$ as $N\to\infty$, but later calculations would be
more complicated wherefore we prefer to work with the signed measures 
$\hat{\mu}^{}_N$ rather than with the positive measures $\hat{\mu}^{+}_N$.

\begin{theorem} \label{t1drt}
  Under the assumptions of Proposition \ref{1drt}, the diffraction spectrum 
  consists, 
  with probabilistic certainty, of a pure point (Bragg) part and an absolutely 
  continuous part, so $\hat{\gamma}^{}_{\omega} = 
        (\hat{\gamma}^{}_{\omega})_{pp} + (\hat{\gamma}^{}_{\omega})_{ac}$.
  If $\alpha=u/v$, the pure point part is 
\begin{equation} \label{point3}
  (\hat{\gamma}^{}_{\omega})_{pp} \; = \; d^2 \cdot
    \begin{cases}
         \delta_0  & \text{if $\alpha\not\in\QQ$}, \\
         \sum_{k\in\frac{1}{\xi}\ZZ} \delta_k 
                         & \text{if $\alpha\in\QQ$},
    \end{cases}
\end{equation}
where, if $\alpha\in\QQ$, we set $\alpha=a/b$ with coprime $a,b \in\ZZ$
and define $\xi=u/a=v/b$.  
  
  The absolutely continuous part $(\hat{\gamma}^{}_{\omega})_{ac}$ can be 
  represented by the continuous function
\begin{equation} \label{cont3}
   g(k) \; = \; 
         \frac{d \cdot p q \sin^2(\pi k(u-v))}
         {p \sin^2(\pi ku) + q \sin^2(\pi kv) - pq \sin^2(\pi k(u-v))} \, ,
\end{equation}
  which is well defined for $k(u-v)\not\in\ZZ$. It has a smooth continuation to 
  the excluded points. If $\alpha$ is irrational, this is $g(k)=0$ for 
  $k(u-v)\in\ZZ$ with $k\neq 0$ and
\begin{equation} \label{cont-irrational}
     g(0) \; = \;  \frac{d\cdot pq (u-v)^2}{p u^2 + q v^2 - pq (u-v)^2}  
          \; = \;  d\, \frac{\, pq (u-v)^2}{(p u + q v)^2} \, .
\end{equation}
  {}For $\alpha=a/b\in\QQ$ as above, it is $g(k)=0$ for $k(u-v)\in\ZZ$, but
  $ku\not\in\ZZ$ (or, equivalently, $kv\not\in\ZZ$), and 
\begin{equation} \label{cont-rational}
     g(k) \; = \; d\, \frac{\, pq (a-b)^2}{(p a + q b)^2}
\end{equation}
  for the case that also $ku\in\ZZ$.
\end{theorem}
{\sc Proof}: 
We employ the sequence of measures $\mu^{}_N$ introduced above which converges
weakly to $\gamma^{}_{\omega}$. A direct calculation shows that the tempered
measure $\hat{\mu}^{}_N$ is represented by the analytic function
\begin{equation} \label{formal}
   g^{}_N(k) \; = \; d \, \sum_{n=-N}^{N}
    \left(p e^{-{\rm sgn}(n) 2\pi iku} + 
          q e^{-{\rm sgn}(n) 2\pi ikv}\right)^{|n|}\, .
\end{equation}
If we define $r(k) = p e^{-2\pi iku} + q e^{-2\pi ikv}$, it is clear that this 
is a complex number inside the (closed) unit circle, i.e.\ $r(k)=R e^{i\phi}$
with $0\leq R\leq 1$ and $\phi\in[0,2\pi)$. This results in
\begin{equation}
  g^{}_N(k) \; = \; d \,\left( 1 + 2 \,\sum_{m=1}^N R^m \cos(m\phi) \right)\, ,
\end{equation}
which shows that $g^{}_N(k)$ represents a sequence of signed (or real), but not
necessarily positive, measures. Their limit, however, is positive.

Let us first check where the sequence of functions converges pointwise. We have 
$r(-k)=\overline{r(k)}$. Then, by the triangle inequality,
$|r(k)|\leq p+q = 1$. Also, we have $|r(k)|^2 = 1 - 4pq\sin^2(\pi k(u-v))$, and
thus $|r(k)|^2=1$ if and only if $p=0$, $p=1$, or $k(u-v)\in\ZZ$. We have
excluded the trivial cases $p=0$ and $q=0$ by our assumptions (they correspond
to periodic chains, up to defects of density zero). So, if $k(u-v)\not\in\ZZ$,
the geometric series in (\ref{formal}) actually converges, with limit
\begin{equation}\label{geom-limit}
   g(k) \; = \; d \; \frac{1-|r(k)|^2}{|1-r(k)|^2} \, ,    
\end{equation}
which immediately gives the expression in (\ref{cont3}). In particular, the 
denominator is always different from 0 for $k(u-v)\not\in\ZZ$.

It is not difficult to check that $g(k)$  has a continuation to points $k$ with 
$k(u-v)\in\ZZ$. Consider first the case $\alpha$ irrational. Then, for 
$k\neq 0$,
the denominator of (\ref{cont3}) is never 0, and $g(k)=0$ is the correct
continuation. The case $k=0$ requires twice the application of de l'Hospital's
rule, and gives the value of $g(0)$ of (\ref{cont-irrational}). Next, let
$\alpha=a/b$ with coprime $a,b\in\ZZ$. If $k(u-v)\in\ZZ$, then $ku\in\ZZ$ if
and only if $kv\in\ZZ$. So, if $ku\not\in\ZZ$, we are back to the case where
$g(k)=0$ is the correct continuation, while $ku\in\ZZ$, again with de 
l'Hospital's
rule, gives the extension stated in (\ref{cont-rational}). This in particular
demonstrates that $g(k)$, with the appropriate continuation, is a continuous
function. It is also a positive function, as can easily be checked, and thus
represents an absolutely continuous positive measure.

So, we have shown that $g^{}_N(k) - g(k)$ tends pointwise to 0, as $N\to\infty$,
for all $k$ with $k(u-v)\not\in\ZZ$. The convergence is actually uniform on
each compact interval that does not contain any of the exceptional points, as is 
clear from (\ref{geom-limit}).
The latter form a 1D lattice of spacing $1/(u-v)$, and any singular part of 
$\hat{\gamma}^{}_{\omega}$ must thus be concentrated to this set.
Since the latter is uniformly discrete, the singular part cannot be
singular continuous, but at most consist of point measures. 

We now have to check what happens with $g^{}_N(k)$ for $k(u-v)\in\ZZ$, where 
the sequence of functions does {\em not} converge. First, let $\alpha$ be
irrational, but $k\neq 0$. Then, it is impossible to have $r(k)=1$, because
this would imply $ku\in\ZZ$ and hence $kv\in\ZZ$ -- a contradiction to
$\alpha\not\in\QQ$. But then, with $r(-k)=\overline{r(k)}$ and $|r(k)|\leq 1$, 
we get 
\begin{eqnarray} \label{converg}
   \frac{1}{d} \, |g^{}_N(k)| 
   & \leq & 1 + 2 \, \left| \sum_{n=1}^{N} r(k)^n  \right|
               \; = \; 1 + 2 \, \frac{|r(k)| \, |1-r(k)^{N}|}{|1-r(k)|}
               \nonumber \\
   & \leq & 1 + 2 \, \frac{1 + |r(k)|^N}{|1-r(k)|}
               \; \leq \; 1 + \frac{4}{|1-r(k)|} \\
   & = & 1 + 
    \frac{2}
    {\sqrt{p \sin^2(\pi ku) + q \sin^2(\pi kv)
     - p q \sin^2(\pi k(u-v))} } \nonumber
\end{eqnarray}
which, for each fixed $k\neq 0$, has a denominator $\neq 0$. So, the sequence 
$g^{}_N(k)$, and hence (\ref{formal}), stays bounded in this case, even 
though it does not converge. With the previous result on $g$, this means that 
the sequence $g^{}_N$ is uniformly bounded on each closed interval that does not
include $0$. Hence, $\hat{\gamma}^{}_{\omega}$ cannot be singular at the
exceptional points $k$ with $k(u-v)\in\ZZ$ unless $k=0$. The analogous argument 
applies if $\alpha=a/b\in\QQ$, as long as $ku\not\in\ZZ$: the sequence $g^{}_N$ 
is uniformly bounded on each closed interval that does not include any of the 
exceptional points with $ku\in\ZZ$, which we will call singular from now on.

So, the remaining cases are $k=0$ for $\alpha$ irrational resp.\
$ku\in\ZZ$ for $\alpha$ rational. For such singular $k$, (\ref{formal}) gives
\begin{equation}
    g^{}_N(k) \; = \; d (2N+1) \, ,
\end{equation}
and this means that $g^{}_N(k)$ diverges for these $k$, always with the
same rate, and the divergence is proportional to the system size. To make
this precise, consider first $\alpha=a/b$ with coprime $a,b\in\ZZ$. The 
singular points are then $k=\ell a/u=\ell b/v$ for $\ell\in\ZZ$. Define
$L(N)=\lfloor N(p a + q b)\rfloor$, where $\lfloor x\rfloor$ denotes the 
integer part of a positive $x$, and set
\begin{equation}
    h^{}_N(k) \; = \; d^2 \xi \sum_{m=-L(N)}^{L(N)} e^{-2\pi i k m\xi}
\end{equation}
with $\xi=u/a=v/b$. Clearly, $h^{}_N(k) = d (2N+1) + {\cal O}(1)$ for all 
singular $k$. The form of $h^{}_N(k)$ is fixed by the requirement that both 
the height and the width of the finite approximations to the point measures at 
the singular points equals that of $g^{}_N(k)$, up to lower order terms. 
One can now show (though we will skip the details here) that 
$g^{}_N(k) - h^{}_N(k)$ is bounded on each compact interval even if 
it does contain singular values of $k$. On the other hand, we know from 
(\ref{exp-sum}) that
\begin{equation}
    \lim_{N\to\infty} h^{}_N(k) \; = \;
        d^2 \xi \sum_{x\in\xi\ZZ} e^{-2\pi i k x} \; = \;
         d^2 \sum_{y\in\frac{1}{\xi}\ZZ} \delta_y \, .
\end{equation}
This then proves that $g^{}_N(k)$ converges to a point measure concentrated
on $\frac{1}{\xi}\ZZ$ plus the $ac$ measure derived above. Similarly, one
deals with the case $\alpha$ irrational, but $k=0$ (e.g.\ by taking a suitable
sequence of rational cases while letting $\xi\to 0$).

So, for all singular $k$, the structure factor converges to $d^2$, 
i.e.\ $(\hat{\gamma}^{}_{\omega})_{pp} (\{k\})\; = \; d^2$, and we obtain the
result given in Eq.~(\ref{point3}).

Consequently, the positive measure $\hat{\gamma}^{}_{\omega}$ has the 
decomposition claimed, and the absolutely continuous part can be represented 
by the continuous
function $g(k)$ (with the appropriate continuation to all $k$). This is the
Radon-Nikodym density \cite[Thm.\ I.19]{RS} which is uniquely determined almost 
everywhere. \qed \smallskip 

Let us mention that the function $g(k)$ appears simpler than it is --- if one
tries a selection of different parameters and produces some plots, one quickly
realizes that it actually shows some ``spiky'' structure (though it is smooth),
and the way it does depends rather critically on the nature of $\alpha=u/v$.
For example, if $\alpha=\tau$ (the golden ratio), a rather regular pattern
emerges, and localized bell-shaped needles of increasing height appear at
sequences of positions that scale with $\tau$. This is reminiscent of what
happens in perfect Fibonacci model sets. If, however, $\alpha$ is transcendental
(e.g.\ $\pi$), much more pronounced needles appear, but at rather irregular
positions. This is a clear consequence of how these numbers can be approximated
by rationals, and an analogous phenomenon is well known in the approximation
of irrational numbers by finite continued fractions \cite[Ch.\ VII]{Cassels}.

We have discussed the case of a binary random tiling in detail, to make the
structure as transparent as possible. It is clear that one can treat, with the
same methods, also the case of a random tiling with $n$ tiles of length
$(u^{}_1,\ldots,u^{}_n)=\bs{u}$, $u^{}_i>0$, and attached frequencies
$(p^{}_1,\ldots,p^{}_n)=\bs{p}$, $p^{}_i>0$, $\sum_{i=1}^{n}p^{}_i=1$. 
Let us state the results in an informal way, as they are extremely parallel
to what we discussed above.
Viewing this system again as a Bernoulli system in the symbols $u^{}_i$ reveals
that the mean free path between two consecutive points of $\Lambda$ is, almost 
surely, given by $\bs{p}\cdot\bs{u}$, and the density is then 
$d=1/(\bs{p}\cdot\bs{u})$. To simplify the following formulas, it is 
advantageous to adopt standard multi-index notation. 
So, $\bs{m}=(m^{}_1,\ldots,m^{}_n)$ is a vector of non-negative 
integers, $|\bs{m}|^{}_1 = m^{}_1+\cdots +m^{}_n$ its 1-norm, and
$\bs{p}^{\bs{m}}=p_1^{m^{}_1}\cdot\ldots\cdot p_n^{m^{}_n}$.
Also, we shall need the multinomial coefficient
\begin{equation}
    \binom{N}{\bs{m}} \; = \; 
    \frac{N!}{m^{}_1! m^{}_2! \cdot\ldots\cdot m^{}_n!}
\end{equation}
where $N=|\bs{m}|^{}_1$.

Let $\Lambda$ be the vertex set of a random tiling of this kind.
The natural autocorrelation of $\omega=\omega^{}_{\Lambda}$
exists with probabilistic certainty, and has the form 
$\gamma^{}_{\omega} = \sum_{z\in\Delta} \nu(z)\delta_z$ where
$\Delta = \{ \pm z \mid z = \bs{m}\cdot\bs{u} \}$
and the autocorrelation coefficient is given by
\begin{equation}
 \nu(z) \; = \; d \sum_{\bs{m}\cdot\bs{u}=z} 
        \binom{|\bs{m}|^{}_1}{\bs{m}} \bs{p}^{\bs{m}} \, .
\end{equation}
This is the previous result with the binomial structure replaced by
a multinomial one. In particular, we also get an analogue of 
Eq.~(\ref{amaz1}), namely
\begin{equation}
   \sum_{|\bs{m}|^{}_1=N} \binom{N}{\bs{m}} \bs{p}^{\bs{m}} \,
      (\bs{m}\cdot\bs{u}) \; = \; N \, (\bs{p}\cdot\bs{u}) \, .
\end{equation}

The autocorrelation measure $\gamma^{}_{\omega}$ can again be approximated
by a weakly converging sequence of measures $\mu^{}_N$, and their Fourier
transform now reads
\begin{equation}
  \hat{\mu}^{}_N \; = \; d\cdot \sum_{m=-N}^{N}
    \left( \sum_{j=1}^{n} p^{}_j e^{-{\rm sgn}(m) 2\pi iku^{}_j} 
           \right)^{|m|} \, .
\end{equation}
The analysis of pointwise convergence then reveals once again that the 
diffraction spectrum consists of a pure point part and an absolutely 
continuous part.

The absolutely continuous part of the diffraction, 
$(\hat{\gamma}^{}_{\omega})_{ac}$,
is represented by the continuous Radon-Nikodym density
\begin{equation}
  g(k) \; = \; \frac{d \cdot \sum_{j<\ell} \, p^{}_{j} p^{}_{\ell}
       \sin^2(\pi k(u^{}_j - u^{}_{\ell}))}
    {\sum_j \, p^{}_j \sin^2(\pi k u^{}_j) - \sum_{j<\ell} \, 
        p^{}_{j} p^{}_{\ell}
        \sin^2(\pi k(u^{}_j - u^{}_{\ell}))} \, ,
\end{equation}
which is well defined as long as not all $k(u^{}_j - u^{}_{\ell})$ are integer.
If they are, the continuation is to $g(k)=0$ if $k\bs{u}\not\in\ZZ^n$ and 
otherwise to
\begin{equation}
   g(k) \; = \; 
     \frac{d \cdot \sum_{j<\ell}\, p^{}_j p^{}_{\ell} (u^{}_j -u^{}_{\ell})^2}
        {\sum_j \, p^{}_j u^2_j \, - \, 
         \sum_{j<\ell}\, p^{}_j p^{}_{\ell} (u^{}_j -u^{}_{\ell})^2}
   \; = \; \frac{d \cdot \sum_{j<\ell}\, p^{}_j p^{}_{\ell} 
           (u^{}_j -u^{}_{\ell})^2}
        {(\bs{p}\cdot\bs{u})^2} \, .
\end{equation}

The Bragg part of the diffraction is determined by the condition that, whenever
$k\bs{u}\in\ZZ^n$, $k$ is a pure point, and results in a contribution of
$d^2 \delta_k$ to $(\hat{\gamma}^{}_{\omega})_{pp}$. Thus, if $\bs{u} =
\xi (a^{}_1,\ldots,a^{}_n)$ with $a^{}_i\in\ZZ$ and
$\gcd(a^{}_1,\ldots,a^{}_n)=1$, the condition $k\bs{u}\in\ZZ^n$ is equivalent
to $k\xi\in\ZZ$, and we get a 1D lattice Dirac comb,
\begin{equation}
   (\hat{\gamma}^{}_{\omega})_{pp} \; = \; d^2 \cdot
        \sum_{k\in\frac{1}{\xi}\ZZ} \delta_k \, .
\end{equation}
If, however, at least one quotient $u^{}_i/u^{}_j$ is irrational, we only get
the trivial point part, $d^2 \delta_0$, while all other peaks are extinct.

One reason for this rather detailed discussion will become apparent shortly
when we use this to describe a class of very simple stochastic tilings in higher
dimension.

\subsection*{Intermezzo: Stochastic product tilings}

With the use of 1D random tilings one can construct a particularly simple
class of stochastic tilings in higher dimension, by simply taking one 1D random
tiling per Cartesian direction and considering the space filling by cuboids
obtained that way. The prototiles are thus cuboids whose edges in direction
$j$ is any of the possible lengths of the $j$th 1D random tiling used. Since
the diffraction theory of these objects is essentially an exercise in direct
products, but nevertheless quite useful and instructive, we describe the
result in an informal way. Note, however, that the entropy density of these
tilings is zero if $D>1$, so that they are no ordinary random tilings in the 
sense of \cite{Henley} or \cite{Richard}.

Consider $D$ different 1D random tilings, and the corresponding point sets 
$\Lambda_i$,
$1\leq i\leq D$, characterized by vectors of possible tile lengths 
$\bs{u}^{(i)}$
and frequency vectors $\bs{p}^{(i)}$. The total number of tiles in each case
may be different, and is given by $n_i$. Let us now consider the Cartesian 
product
\begin{equation}
  \Lambda \; = \; \Lambda^{}_1\times\ldots\times\Lambda^{}_D
          \; = \; \{ (x^{}_1,\ldots,x^{}_D)\mid x^{}_i\in\Lambda_i \} \, ,
\end{equation}
which is the vertex set of a stochastic tiling in $D$ dimensions whose 
prototiles
are the $n^{}_1\cdot\ldots\cdot n^{}_D$ cuboids obtained as Cartesian products
of the intervals $u^{(i)}_j$ with $1\leq i\leq D$ and $1\leq j\leq n_i$.
The sets $\Lambda$ are thus all of finite local complexity, and we have
\begin{equation}
  \Delta \; = \; \Lambda-\Lambda \; = \;
     \{ \bs{z}=(z^{}_1,\ldots,z^{}_D)\mid z^{}_i\in\Delta_i \}
\end{equation}
with probability one, where $\Delta_i=\Lambda_i - \Lambda_i$. Also, the density
of $\Lambda$ exists again with probabilistic certainty, and is given by
$d=d^{}_1\cdot\ldots\cdot d^{}_D$ where 
$d_i = (\bs{p}^{(i)}\cdot\bs{u}^{(i)})^{-1}$
as derived in the previous Section.

The product structure of $\Lambda$ also implies that 
$\omega=\sum_{\bs{x}\in\Lambda}\delta_{\bs{x}}$ is a product measure (or,
which is equivalent in this case, the tensor product of distributions, see
\cite[Chap.\ IV]{Schwartz}), i.e.\ we have
\begin{equation}
   \omega \; = \; \prod_{i=1}^D \omega^{(i)}
          \; = \; \prod_{i=1}^D 
          \left( \sum_{x_i\in\Lambda_i}\delta^{(i)}_{x_i} \right) \, ,
\end{equation}
where $\delta^{(i)}$ is meant as a 1D Dirac measure acting in the space along 
the $i$th coordinate.
It also follows, with probabilistic certainty, that the autocorrelation
$\gamma^{}_{\omega}$ exists and is a product measure, too.
To prove the existence of the corresponding coefficients $\nu(\bs{z})$, 
it is easiest to take
averages over cubes rather then balls, i.e.\ to use the definition of 
\cite{Hof}. Here, this gives the same limit as our definition of the natural 
autocorrelation.
One obtains $\nu(\bs{z}) = \prod_{i=1}^D \nu^{(i)}(z_i)$, where $\nu^{(i)}(z_i)$
is the coefficient of the autocorrelation attached to $\Lambda_i$. So we have
\begin{equation}
   \gamma^{}_{\omega} \; = \; \prod_{i=1}^D
     \left( \sum_{z_i\in\Lambda_i} \nu^{(i)}(z_i)\delta^{(i)}_{z_i} \right)\, .
\end{equation}

{}Finally, let us consider the diffraction spectrum $\hat{\gamma}^{}_{\omega}$.
Since it is the Fourier transform of a product measure, it is a product measure
itself \cite[Thm.\ XIV]{Schwartz}, and we thus obtain
\begin{equation}
   \hat{\gamma}^{}_{\omega} \; = \; \prod_{i=1}^D \hat{\gamma}^{}_{\omega^{(i)}}
    \; = \; \prod_{i=1}^D
    \left( \, (\hat{\gamma}^{}_{\omega^{(i)}})^{}_{pp} +
           (\hat{\gamma}^{}_{\omega^{(i)}})^{}_{ac} \, \right) \, ,
\end{equation}
where the $\hat{\gamma}^{}_{\omega^{(i)}}$ are determined through Theorem 
\ref{t1drt}. In particular, the absolutely continuous (pure point) part of 
$\hat{\gamma}^{}_{\omega}$ is precisely the product of the $ac$ ($pp$) parts
of the $\hat{\gamma}^{}_{\omega^{(i)}}$, while all other combinations result
in singular continuous components --- though the meaning of this will require
some thought. The $sc$ property can be seen from the fact that terms in
the expanded product which contain at least one component of each kind are
concentrated to a support of vanishing Lebesgue measure, but contain 
no pure points themselves --- whence they must be singular continuous relative
to Lebesgue measure.
This also agrees with the common, intuitive scaling picture: a term with $m$
$ac$-components and $D-m$ $pp$-components would show intensities that stem from
amplitudes (or Fourier-Bohr coefficients) which show a finite-size scaling with 
$L^{m/2} L^{D-m} = (L^D)^{\beta}$ where $L$ is the linear system extension and
\begin{equation}
    \frac{1}{2} \; \leq \; \beta=1-\frac{m}{2D} \; \leq \; 1 \, .
\end{equation}
Here, $\beta=\frac{1}{2}$ and $\beta=1$ correspond to the cases of $ac$ and $pp$
part, respectively, compare the discussion in \cite[Sec.\ 6]{Hof} and
\cite{Hof-scaling}.

However, this notion of singular continuity is to be taken with a grain of
salt. All we have constructed here are product measures, and such objects
would perhaps not qualify to be singular continuous in a ``generic'' sense.
They do show up in liquid crystals though, compare the discussion on the nature 
of their long-range order in \cite{HL}.
In particular, they appear in Danzer's aperiodic prototile
in 3-space \cite[Sec.\ 4]{Danzer}, which can be seen as a toy
model of a smectic C$^*$ liquid crystal. Nevertheless,
these simple product tilings show that one has to expect a larger variety of
spectral types in higher dimensions, and that already in planar cases the
appearance of singular continuous components should be typical. Also, one can
easily construct examples with all three spectral types present. More specific
and genuine random tilings, however, might bypass this, as we shall see in
the next Section, although that should not be considered generic.


\subsection*{Two-dimensional random tilings}

Let us now move to planar systems, where we will mainly consider two 
illustrative examples, namely the classical random tilings consisting of 
dominoes and lozenges. 
Because of their symmetries, we call them crystallographic random tilings.
Though still only two-dimensional, they are of
practical relevance because of the existence of so-called T-phases (see 
\cite{Baake} and references therein) which are
irregular planar layers stacked periodically in the third direction due to a
very anisotropic growth mechanism. It is thus appropriate to investigate the
diffraction spectrum of a single layer obtaining then the complete spectrum 
 once again as a product measure, compare the previous Section.

Unfortunately, already the treatment of planar systems is a lot more involved
than in the 1D case. Although we will have to deal ``only'' with the action of 
$\ZZ^2$,
we cannot directly apply standard results of ergodic theory as above, because 
we first have to establish the ergodicity of the measures involved.
Even for the two simple systems we shall discuss below, the
rigorous classification of invariant measures is only in its infancy, see
\cite{BP} and the discussion in \cite{Ken1}. Fortunately, the investigation
of invariant equilibrium states, which form a subclass, is well developed 
\cite{Georgii,Israel,Simon}. If combined with certain results of statistical
mechanics \cite{Ruelle}, this allows for a determination of extremal
states, which will be unique in our examples. 
They are ergodic and thus admit the application of Birkhoff's pointwise
ergodic theorem in its version for $\ZZ^2$-action, see e.g.\
\cite[Thm.\ 2.1.5]{Keller}. 
Let us summarize the key features in a way adapted to our later examples.

\subsubsection*{Preliminaries}

A {\em tiling} $\omega$ of a region\footnote{We tacitly assume that any such 
region is sufficiently nice, i.e.\ it should be compact, measurable and simply 
connected.} $\Lambda\subset\RR^d$ (with positive volume ${\rm vol}(\Lambda)$) is a 
countable covering of $\Lambda$ by tiles, i.e.\ by bounded closed sets 
homeomorphic to balls, having pairwise 
disjoint interiors and non-vanishing overlap with the region $\Lambda$. 
In our case, the tiles are translates of finitely many prototiles. 
We deal with free boundary conditions in the sense that the tiles may protrude 
beyond the boundary. Thus, 
the boundary of $\Lambda$ does not impose any restrictions of the kind known 
from fixed boundary conditions or exact fillings of given patches, compare 
\cite{Ken2}. Two coverings of the same region $\Lambda$ are called 
{\em equivalent} if they are translates of one another. For further conceptual 
details, we refer to \cite{Richard}.

The examples discussed below belong to the class of polyomino tilings, where 
the prototiles are combinations of several elementary cells of a given periodic 
graph $G$. These tilings can be described as polymer models \cite{Richard}, and 
as dimer models in our case.
\begin{figure}[ht]
\centerline{\epsfysize=3cm \epsfbox{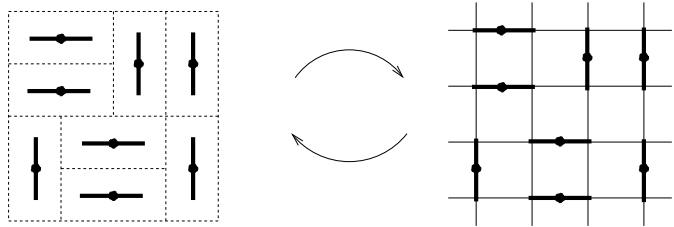}}
\caption{\label{dual}Representation of a random tiling on the dual cell 
complex.}
\end{figure}

A {\em dimer} is a diatomic molecule occupying two connected sites of a graph. 
A graph is close-packed if all sites are occupied precisely once.
In Figure~\ref{dual}, the one-to-one correspondence between the tiling on a 
(periodic) graph and its close-packed dimer configuration on the dual cell 
complex (the so-called
Delone complex) is illustrated for the domino tiling. The scatterers (Dirac unit
measures) are placed in the centre of the tiles, resp.\ dimers. 

In the space $\Omega$ of all tilings, two elements are close if they agree on a 
large neighbourhood of the origin. $\Omega$ is compact in this topology. 
The group $\ZZ^2$ of translations acts continuously on $\Omega$ (in an 
appropriate parametrization) 
because of the periodicity of the underlying graph. 

We use a grand-canonical setup where we assign equal (zero) interaction energy 
and a finite {\em chemical potential} $\mu_i$ or activity $z_i=e^{\mu_i}$ to 
each of the $M$ different prototiles (setting the inverse temperature 
$\beta=1$). For fixed prototile 
numbers $n_1,\dots,n_M$, let us denote the number of nonequivalent 
$\Lambda$-patches that use $n_i$ prototiles of type $i$ by
$g_{\Lambda}(n_1,\dots,n_M)$. The {\em grand-canonical partition function} is 
given by the following configuration generating function 
($\mu=(\mu_1^{},\dots,\mu_M^{})$)
\begin{equation}
   \mathcal{Z}_{\Lambda}^{}(\mu) \; = \; 
        \sum_{n^{}_1,\dots,n^{}_M}g^{}_{\Lambda}(n_1,\dots,n_M)\, 
               z_1^{n^{}_1} \cdot\ldots\cdot z_M^{n^{}_M}.
\end{equation}
To adapt the usual language of statistical mechanics, let 
$\omega\in\Omega_{\Lambda}$ be a tiling of a 
finite region $\Lambda$ which is now positioned relative to a fixed
lattice, $\ZZ^2$ say. An {\em interaction} is the translation invariant 
assignment of 
a continuous function $\Phi(\omega)\in C_{\omega}$ to every  $\omega$, so that 
a shift of $\Lambda$ by $a\in\ZZ^2$ results in
$\Phi(\omega+a)=(\tau_a \Phi)(\omega)$, where $\tau_a$ denotes the corresponding
shift (cf.\ \cite{Ruelle,Simon}). To avoid repeated counting of contributions
to the interaction energy, $\Phi$ represents only the basic interaction of 
tiles in $\omega$ that are not already included in the interaction of 
subsystems. For $\omega\subset\Omega^{}_{\Lambda}$, we now define the 
Hamiltonian $H_{\Lambda}^{\Phi}(\omega)$ by 
\begin{equation}
   H_{\Lambda}^{\Phi}(\omega) \; = \; 
                \sum_{\omega' < \, \omega}\Phi(\omega') \, ,
\end{equation}
where the sum runs over all sub-patches of $\omega$. This is well defined due
to the restrictions on $\Phi$ mentioned before. Let $\mathcal{B}$ be the Banach 
space (with norm $\|.\|_{\infty}$) of interactions $\Phi$ subject to the 
restriction
\begin{equation}
     |||\Phi||| \; := \; \sum_{\omega\ni 0}
                \frac{\|\Phi(\omega)\|_{\infty}}{|\omega|}
                \; < \; \infty \, ,
\end{equation}
where the sum is over all tilings covering the origin and $|\omega|$ denotes 
the number of tiles (cf.\ also \cite[App.~B]{Israel}). 
Let us now, for simplicity, assume that each tiling $\omega$ of $\Lambda$ has 
the same total number of tiles, $|\omega|=N^{}_{\Lambda}$. The {\em pressure\/} 
(negative grand-canonical potential) per tile is then given by 
\begin{equation}
   p^{}_{\Lambda}(\Phi) \; = \; \frac{1}{N^{}_{\Lambda}}
        \sum_{\omega\in\Omega_{\Lambda}}e^{-H_{\Lambda}^{\Phi}(\omega)}
        \; = \; \frac{1}{N^{}_{\Lambda}}\log\mathcal{Z}_{\Lambda}^{}(\Phi) \, .
\end{equation}
If the extra assumption is not fulfilled, $N^{}_{\Lambda}$ is the average 
number of tiles. This is a reasonable definition 
 as long as we take the limit $\Lambda\to\infty$ in the sense of van 
Hove, see \cite{Ruelle} for details on this concept. But then, for 
$\Phi\in\mathcal{B}$,
the thermodynamic limit $p(\Phi)=\lim_{\Lambda\to\infty}p_{\Lambda}(\Phi)$ 
exists and is a convex function of $\Phi$, compare \cite{Israel,Ruelle,Simon}. 
The interaction in our models is simply the self-energy ($-\mu_i$ for 
prototiles of type $i$) and thus included in $\mathcal{B}$. 

A {\em state} $\nu$ of the infinite system is a Borel probability measure on 
$\Omega$. A state $\nu$ is called {\em invariant} or 
{\em translation invariant} if it is 
invariant under the action of $\ZZ^2$. The set of all invariant measures, 
$\mathcal{M}^I$, is compact and forms a simplex.

Following Ruelle's presentation \cite[Ch.~7.3]{Ruelle}, let $D$ be the set of 
all $\Phi\in\mathcal{B}$ such that the graph of $p$ 
has a unique tangent plane at the 
point $(\Phi,p(\Phi))$. If $\Phi\in\mathcal{B}$, there exists a unique linear 
functional $\alpha^{\Phi}$ in the dual $\mathcal{B}^*$ of $\mathcal{B}$ 
such that 
\begin{equation} \label{alpha}
    p(\Phi+\Psi) \; \geq \; p(\Phi)-\alpha^{\Phi}(\Psi) \, .
\end{equation}
The {\em mean densities} 
$\rho_i^{\Lambda}(\mu)=\frac{\langle n_i\rangle_{\Lambda}}{{\rm vol}(\Lambda)}$,
$i=1,\dots,M$, of the different prototiles in the ensemble can be computed as 
functional derivative of $p_{\Lambda}(\mu)$ with respect to $\mu$
($\langle\,.\,\rangle_{\Lambda}$ denotes the finite-size average for given 
chemical
potentials $\mu_1,\dots,\mu_M$). For $\mu\in\mathcal{B}$ and $\Lambda\to\infty$ 
in the sense of van Hove, we have\footnote{Usually one defines 
$A_{\Phi}(\omega)=\sum_{\omega' < \, \omega}
\frac{\Phi(\omega')}{|\omega'|}\in\mathcal{B}^*$.
Restricting ourselves to the tiling models where the chemical potentials are 
the only interactions, we may identify $A_{\mu}$ and $\mu$.}
\begin{equation}
    \lim_{\Lambda\to\infty}\sum_{i=1}^M\rho_i^{\Lambda}(\mu)\tilde{\mu}_i
          \; = \; \alpha^{\mu}(\tilde{\mu})
\end{equation}
for all $\tilde{\mu}\in\mathcal{B}$, and
\begin{equation}
     \lim_{\Lambda\to\infty}\rho_i^{\Lambda} \; = \; \rho_i^{}
           \; = \; \frac{\partial p(\mu_1,\dots,\mu_M)}{\partial\mu_i} \, ,
\end{equation}
with $\sum_{i=1}^M \rho_i=1$. Since the chemical potentials and also 
the conjugate (mean) densities do not form an independent set of macroscopical
parameters, we may choose an independent subset 
$\rho_1^{},\dots,\rho_k^{}$ by setting
$\mu_{k+1}^{}=\dots=\mu_M^{}=0$. This normalization of $p$ leaves the densities 
invariant.

The {\em entropy} per tile of a finite region tiling is defined as
\begin{equation}
    s(\nu^{}_{\Lambda}) \; = \; 
          -\frac{1}{N^{}_{\Lambda}}\, \sum_{\omega\in\Omega_{\Lambda}} 
           \nu^{}_{\Lambda}(\omega)\log \nu^{}_{\Lambda}(\omega) 
\end{equation}
where $\nu^{}_{\Lambda}$ is the restriction of $\nu$ to $\Omega_{\Lambda}$.
For translation invariant measures and $\mu\in\mathcal{B}$, the infinite volume 
limit exists, giving $\nu^{}_{\Lambda}\to\nu$ for $\Lambda\to\infty$ taken in 
an appropriate way, and the functional $s$ is affine upper semicontinuous 
\cite[Ch.~7.2]{Ruelle}. According to Gibbs' variational principle 
\cite[Ch.~7.4]{Ruelle}, the pressure can  then be calculated as
\begin{equation}
     p(\mu) \; = \; \sup_{\nu\in\mathcal{M}^I}[s(\nu)-\nu(\mu)]\, .
\end{equation}
The measure for which this supremum is attained is called {\em equilibrium 
measure}. We may formulate the weak Gibbs phase rule \cite[Ch.~7.5]{Ruelle}.

\begin{theorem} Let $D\subset\mathcal{B}$ defined as above.
\begin{enumerate} 
  \item If $\mu\in D$, the function $\nu\mapsto s(\nu)-\nu(\mu)$ reaches its 
        maximum $p(\mu)$ at exactly one point $\nu^{\mu}\in\mathcal{M}^I$.
  \item If $\mu\in D$ and $\alpha^{\mu}\in\mathcal{B}^*$ is defined by
        (\ref{alpha}), then, for all $\tilde{\mu}\in\mathcal{B}$,
  \begin{equation}
        \nu^{\mu}(\tilde{\mu}) \; = \; \alpha^{\mu}(\tilde{\mu})
  \end{equation}
        so that $\nu^{\mu}$ is the infinite volume equilibrium state 
        corresponding to the chemical potential $\mu$.
  \item If $\mu\in D$, $\nu^{\mu}$ is a $\ZZ^2$-ergodic state and may thus be
        interpreted as pure thermodynamic phase. \qed
\end{enumerate}     
\end{theorem}

In what follows, we calculate the ensemble average of the correlations.
Since the diffraction image is taken from a single member of the ensemble, the 
above theorem ensures that the {\em typical} member is self-averaging as long 
as the pressure is differentiable (no first order phase transition).

Calculating the diffraction of our models consists essentially in calculating 
the corresponding dimers autocorrelation which we will base upon previous work 
of Fisher and Stephenson \cite{FS} and of Kenyon \cite{Ken1}.

Kasteleyn \cite{Kast2} has shown that for any finite planar graph with even 
number of sites, and also for any periodic graph with a fundamental cell of an 
even number of sites, a Pfaffian\footnote{A Pfaffian is basically the square 
root of the determinant of an even antisymmetric matrix, see 
\cite[App.\ E]{Thompson} or \cite[Ch.\ IV.2]{McCoy} for an introduction.}   
can be constructed which is equal to the dimer generating function. For this, 
one has to orientate the graph in such a way that every configuration is 
counted with the correct sign. In addition, every bond is weighted with the 
corresponding dimer activity $z_i$. The configuration function is then given 
by the Pfaffian of the activity-weighted adjacency matrix $\bs{A}$. Although 
Kasteleyn's proof applies to arbitrary graphs, the calculations simplify 
considerably when restricted to periodic simply connected graphs
as in our case. If we define occupation variables for a bond between sites 
${\bs k}$ and ${\bs k}'$
\begin{equation}
    \eta^{}_{\bs{kk}'} \; = \;
           \begin{cases}
               1, \qquad\text{bond (${\bs k},{\bs k}'$) occupied,} \\
               0, \qquad\text{otherwise,}
           \end{cases}
\end{equation}
we can state \cite{FS}
\begin{prop}
  Let $G$ be an infinite simply connected periodic graph with close-packed dimer
  configuration where each dimer orientation has density $\rho_i^{}>0$. 
  Let $\bs{A}$ be the invertible weighted adjacency matrix. If the dimer 
  autocorrelation (joint occupation probability) exists, it is given by 
\begin{eqnarray} \label{auto2d}
    P_{\alpha\beta} & = & \langle \eta^{}_{\bs{k}_{\alpha}^{}\bs{k}'_{\alpha}}
                    \eta^{}_{\bs{k}_{\beta}^{}\bs{k}'_{\beta}}\rangle \\
      & = & \langle\eta^{}_{\bs{k}_{\alpha}^{}\bs{k}'_{\alpha}}\rangle\langle
              \eta^{}_{\bs{k}_{\beta}^{}\bs{k}'_{\beta}} \rangle-
        A_{\bs{k}_{\alpha}^{}\bs{k}'_{\alpha}}
        A_{\bs{k}_{\beta}^{}\bs{k}'_{\beta}}
        (A^{-1}_{\bs{k}_{\alpha}^{}\bs{k}_{\beta}^{}} A^{-1}_{\bs{k}'_{\alpha}
        \bs{k}'_{\beta}} -A^{-1}_{\bs{k}_{\alpha}^{}\bs{k}'_{\beta}}
        A^{-1}_{\bs{k}'_{\alpha}\bs{k}_{\beta}^{}})\, .  \nonumber 
\end{eqnarray}  
\end{prop}
{\sc Proof}: Let $\bar{\eta}_{\bs{kk}'}^{}=1-\eta^{}_{\bs{kk}'}$. Obviously,
\begin{equation}\label{auto2dgen}
    \langle \eta^{}_{\bs{k}_{\alpha}^{}\bs{k}'_{\alpha}}
            \eta^{}_{\bs{k}_{\beta}^{}
    \bs{k}'_{\beta}}\rangle 
    \; = \; \langle\eta^{}_{\bs{k}_{\alpha}^{}\bs{k}'_{\alpha}}\rangle\langle
         \eta^{}_{\bs{k}_{\beta}^{}\bs{k}'_{\beta}} \rangle + \langle
         \bar{\eta}_{\bs{k}_{\alpha}^{}\bs{k}'_{\alpha}}
         \bar{\eta}_{\bs{k}_{\beta}^{}\bs{k}'_{\beta}}\rangle-\langle
         \bar{\eta}_{\bs{k}_{\alpha}^{}\bs{k}'_{\alpha}}\rangle\langle
         \bar{\eta}_{\bs{k}_{\beta}^{}\bs{k}'_{\beta}} \rangle\, .
\end{equation}
The first term on the RHS of (\ref{auto2dgen}) depends only on the 
densities $\rho_{\alpha}^{}$ resp.\ $\rho_{\beta}^{}$ of the dimers that can 
occupy the bond $(\bs{k}_{\alpha}^{}\bs{k}'_{\alpha})$ resp.\ 
$(\bs{k}_{\beta}^{}\bs{k}'_{\beta})$.
In the case of $\bs{k}_{\alpha}^{}$ and $\bs{k}'_{\alpha}$ being connected by 
a bond of type $i$ and only one bond of this type leading to each site, this 
would result in
$\langle \eta^{}_{\bs{k}_{\alpha}^{}\bs{k}'_{\alpha}}\rangle=\rho_{i}$, because 
we normalize with respect to the total number of dimers (and not to the number 
of sites or bonds as in \cite{FS}). It remains to prove the equivalence of the 
second terms of (\ref{auto2d}) and (\ref{auto2dgen}). This was shown in 
\cite{FS}, so we will just give an outline here.

Here, $\langle\bar{\eta}_{\bs{k}_{\alpha}^{}\bs{k}'_{\alpha}}\rangle$ is the 
(weighted) sum of all dimer configurations 
where the bond $(\bs{k}_{\alpha}^{}\bs{k}'_{\alpha})$ is {\em not} occupied, 
divided by the total (weighted) sum of configurations
$\mathcal{Z}$. Since $\mathcal{Z}=\text{Pf}(\bs{A})$, we define
$\text{Pf}(\bs{\tilde{A}})=\text{Pf}(\bs{A}+\bs{E})$ as the ``perturbed'' 
Pfaffian counting precisely all configurations where 
$(\bs{k}_{\alpha}^{}\bs{k}'_{\alpha})$ is 
not occupied, i.e.\ all elements of $\bs{E}$ are zero except 
$E_{\bs{k}_{\alpha}^{}\bs{k}'_{\alpha}}=-E_{\bs{k}'_{\alpha}\bs{k}_{\alpha}^{}}=
-A_{\bs{k}_{\alpha}^{}\bs{k}'_{\alpha}}$. Note that $\bs{A},\bs{\tilde{A}}$ 
and $\bs{E}$ are skew-symmetric matrices. So 
$\langle\bar{\eta}_{\bs{k}_{\alpha}^{}\bs{k}'_{\alpha}}\rangle
=\text{Pf}(\bs{\tilde{A}})/\mathcal{Z}
=\text{Pf}(\II+\bs{A}^{-1}\bs{E})=\text{Pf}(\bs{E})\;
\text{Pf} (\bs{E}^{-1}+\bs{A}^{-1})$, where the last equality holds only if 
$\bs{E}$ is 
invertible, $\II$ denotes the unit matrix of appropriate dimension. The same 
applies to $\langle \bar{\eta}_{\bs{k}_{\alpha}^{}\bs{k}'_{\alpha}}
\bar{\eta}_{\bs{k}_{\beta}^{} \bs{k}'_{\beta}}\rangle$ but with four 
nonvanishing elements of $\bs{E}$. The evaluation of the Pfaffian reduces to 
the calculation of a small determinant and yields the desired result. 
\qed \smallskip

The calculation of $\mathcal{Z}$ and ${\bs A}^{-1}$ can be simplified 
considerably by imposing periodic boundary conditions. Usually, for the 
partition function for a 
toroidal graph, one needs four determinants differing from that with free 
boundary only in exactly these boundary elements \cite{Kast}. But in the 
infinite volume limit, these modifications do not change the value of a 
determinant (given by the product of the eigenvalues) as can be derived from 
the following result of Ledermann 
\cite{Led}
\begin{lemma} 
   If in a Hermitian matrix the elements of $r$ rows and their
   corresponding columns are modified in any way whatever, provided the matrix 
   remains Hermitian, then the number of eigenvalues that lie in any given 
   interval cannot increase or decrease by more than $2r$. \qed
\end{lemma}
If $N=mn$ is the number of sites or elementary cells, the number of eigenvalues 
per unit length, which is $\mathcal{O}(N)$, changes only by 
$\mathcal{O}(\sqrt{N})$, which is negligible in 
the limit as $N\to\infty$ (for details see also \cite{Montroll}). The same 
argument holds for $\bs{A}^{-1}$.

How can we calculate the elements of $\bs{A}^{-1}$? Since $\bs{A}$ is the 
adjacency matrix of a graph made up of a periodic array of elementary cells 
with toroidal boundary conditions and is therefore cyclic, 
it can be reduced to the diagonal form 
$\bs{\Lambda}=\text{diag}\{\lambda_{\bs j}\}$  
by a Fourier-type similarity transformation with matrix elements $S_{\bs{kk}'}=
(mn)^{-1/2}\exp(2\pi i (k_1^{} k'_1/m +k_2^{} k'_2/n))$. $\bs{A}^{-1}$ is now 
determined by
\begin{equation} \label{diag}
   A^{-1}_{\bs{kk}'} \; = \; 
       \left(\bs{S\Lambda}^{-1}\bs{S}^{-1}\right)_{\bs{kk}'}
       \; = \; \sum_{\bs{j}=(1,1)}^{(m,n)} S_{\bs{kj}}^{}
               \lambda_{\bs{j}}^{-1}S^{\dagger}_{\bs{k}'\bs{j}} \, .
\end{equation}
In the infinite volume limit, the sums approach integrals (Weyl's Lemma), and 
by introducing $\varphi_1^{}=2\pi i j_1^{}/m$ etc. we obtain
\begin{equation} \label{inverse}
     A^{-1}_{\bs{kk}'} \; = \; \frac{1}{4 \pi^2}\int_0^{2 \pi}
         \int_0^{2 \pi}\lambda^{-1}(\varphi_1^{},\varphi_2^{})
         e^{-i(\varphi_1^{}(k'_1-k_1^{})+\varphi_2^{}(k'_2-k_2^{}))}
         d\varphi_1^{} d\varphi_2^{}.
\end{equation}
Let us illustrate this by two examples, see \cite{Hoeffe} for another case.

\subsubsection*{Domino tiling}

A domino is a $2$ by $1$ or $1$ by $2$ rectangle, whose vertices have integer 
coordinates in the plane. A tiling of the plane with dominoes is equivalent to 
a close-packed dimer configuration on the square lattice $\ZZ^2$. 
This model exhibits 
no phase transition. We assume finite, positive activities in order to have 
non-vanishing tile densities. The degenerate case will be treated separately. 
Labelling the sites by $\bs{k}=(k_1,k_2)$, and adopting from the various equivalent
possibilities (see \cite{FS} for details) the technically most convenient 
choice of complex weights, one obtains the weighted adjacency matrix 
\begin{align}
    A(k_1,k_2;k_1+1,k_2) & = -A(k_1+1,k_2;k_1,k_2)=z_1,    \nonumber\\
    A(k_1,k_2;k_1,k_2+1) & = -A(k_1,k_2+1;k_1,k_2)=iz_2,            \\
    A(k_1,k_2;k_1',k_2') & =  A_{\bs{kk}'} = 0, \qquad\text{otherwise}, \nonumber
\end{align}
with eigenvalues
\begin{equation}
   \lambda(\varphi_1^{},\varphi_2^{})
   \; = \; 2i\left(z_1^{} \sin \varphi_1^{}+i z_2^{}\sin\varphi_2^{}\right) \, .
\end{equation}
After inserting in (\ref{inverse}) and separating real and imaginary parts, 
one gets ($\bs{k}'-\bs{k}={\bs r}=(x,y)\in\ZZ^2$)
\begin{equation} \label{dominv}
   A^{-1}_{\bs{kk}'} \; = \; 
       \frac{1}{2 \pi^2}\int_0^{\pi}\int_0^{\pi}\frac{M(x,y|\varphi_1^{},
                \varphi_2^{})}{z_1^2 \sin^2_{}\varphi_1^{}+z_2^2
                \sin^2\varphi_2^{}}d\varphi_1^{} d\varphi_2^{}
\end{equation}
with
\begin{equation}
     M(x,y|\varphi_1^{},\varphi_2^{}) \; = \;
\begin{cases}
       0, & \text{($x,y$ same parity)} \\
      -z_1^{}\sin\varphi_1^{}\sin x\varphi_1^{}\cos y\varphi_2^{},
              &\text{($x$ odd, $y$ even)} \\
      -i z_2^{}\sin\varphi_2^{}\sin y\varphi_2^{}\cos x\varphi_1^{},
              &\text{($y$ odd, $x$ even)} 
\end{cases}
\end{equation}
(compare with \cite{FS}). We introduce the abbreviation for the 
{\em coupling function}
\cite{Ken1}
\begin{equation}
     [x,y] \; = \; A^{-1}(k_1,k_2;k_1',k_2')\, ,
\end{equation}
obeying $[x,y]=-[-x,-y]$. We place the scatterers in the centres of the tiles 
or equivalently of the dimers. With (\ref{auto2d}) and (\ref{auto2dgen}), 
the joint occupation probability of two horizontal dimers with scatterers at 
distance ${\bs r}$ in the centre of an infinite lattice is given by
\begin{equation}
    P_{11}({\bs r}) \; = \; \frac{\rho_1^2}{4} + c^{}_{11}(\bs r) \; = \;
       \frac{\rho_1^2}{4}-z_1^2\left([x,y]^2-[x-1,y][x+1,y]\right) \, ,
\end{equation}
(similarly for $P_{22}$) where only $c^{}_{11}(\bs r)$ 
(resp.\ $c^{}_{22}(\bs r)$) depends on the distance 
${\bs r}$. For a pair of mutually perpendicular dimers, the possible distance
vectors of the scatterers ${\bs r}+{\bs a}$, ${\bs a}=(-1/2,1/2)^t$, are odd 
half-integer. The joint occupation probability is
\begin{align}
   P_{12}\left({\bs r}+{\bs a}\right) 
        &\; = \; P_{21}\left({\bs r}-{\bs a}\right)  \nonumber\\
        &\; = \; \frac{\rho_1^{}\rho_2^{}}{4}-iz_1 z_2
          \left([x,y][x-1,y+1]-[x,y+1][x-1,y]\right) \, .
\end{align}
The non-constant part is $c^{}_{12}(\bs r)=c^{}_{21}(\bs r)$.
Note that either the first or second part of the term in brackets vanishes 
because of the parity of $x$ and $y$. The full autocorrelation for the positions of
the scatterers is thus given by
\begin{equation}\label{domauto}
    \gamma^{}_{\omega} \; = \; \sum_{{\bs r}\in\ZZ^2_{}}
        \left(P_{11}\left({\bs r}\right)+P_{22}\left({\bs r}\right)\right)
        \delta_{{\bs r}} +\delta_{\bs a}^{}*\sum_{{\bs r}\in\ZZ^2_{}}
        \left(P_{12}\left({\bs r}+{\bs a}\right)+P_{21}\left({\bs r}+{\bs a}
        \right)\right)\delta_{{\bs r}} \, .
\end{equation}

\begin{theorem} \label{thm4}
  Under the above assumptions, with $\rho_1 \rho_2>0$, the diffraction 
  spectrum of the domino tiling exists with
  probabilistic certainty and consists of a pure point and an absolutely 
  continuous part, i.e.\ 
  $\hat{\gamma}_{\omega}^{}=(\hat{\gamma}_{\omega}^{})_{pp}^{}+
  (\hat{\gamma}_{\omega}^{})_{ac}^{}$, with
\begin{equation}
   \left(\hat{\gamma}_{\omega}\right)_{pp} \; = \; 
         \frac{1}{4}\sum_{(h,k)\in\ZZ^2}\left(\rho_1+
                          (-1)^{h+k}\rho_2\right)^2 \delta_{(h,k)} \, .
\end{equation}
  In particular, there is no singular continuous part. Furthermore,
  $\hat{\gamma}_{\omega}^{}$ is periodic with lattice of periods
  $\{ (h,k)\in\ZZ^2\mid h+k \mbox{ even}\}$.
\end{theorem}
{\sc Proof}: The point spectrum can be calculated directly by taking the 
Fourier transform of the constant part of (\ref{domauto}). This requires 
Poisson's summation formula and the
convolution theorem, which leads to the phase factor $(-1)^{h+k}$
and to the periodicity claimed. 
The non-constant part of $\gamma_{\omega}$ is determined by 
$c^{}_{11}$, $c^{}_{12}$ and $c^{}_{22}$. 
To establish our claim, we will show that the formal Fourier series
$\sum_{\bs{r}\in\ZZ^2} c^{}_{ij}(\bs r) e^{-2\pi i \bs{k}\cdot\bs{r}}$ actually
converge to $L^1$-functions and thus represent absolutely continuous measures.

The real coefficients $c^{}_{ij}(\bs r)$ are essentially products of the form 
$[x,y]^2$. One integration in (\ref{dominv}) may be performed explicitly. With 
standard asymptotic methods involving the Laplace transform, see the Appendix 
of \cite{Young}, one obtains the asymptotic behaviour in the limit of large $x$ 
and $y$ as
\begin{equation}
  [x,y] \; \sim \;
      \begin{cases}
          -\frac{1}{\pi}\frac{z_2^{} x}{(z_2^{} x)^2+(z_1^{} y)^2},
                 & \text{($x$ odd, $y$ even)}, \\
          -\frac{i}{\pi}\frac{z_1^{} y}{(z_2^{} x)^2+(z_1^{} y)^2},
                 & \text{($y$ odd, $x$ even)}.
\end{cases}
\end{equation}
Thus, we obtain $0\leq c_{ij}(\bs{r})
=\mathcal{O}\left(\frac{1}{(x^2+y^2+1)^2} \right)<\infty$, 
since $x$ and $y$ have different parity. Note that the implied constant still 
depends on $z^{}_1$ and $z^{}_2$. Now, e.g.\
by referring to Eqs. (6.1.126) and (6.1.32) of \cite{Hansen}, one can see that
$\sum_{y=0}^{\infty}\sum_{x=0}^{\infty}\frac{1}{(x^2+y^2+1)^2}$ converges
(it actually is even less than 2 in value). Consequently,  
by Cauchy's double series theorem, 
$\sum_{\bs{r}\in\ZZ^2_{}}\left(c_{ij}(\bs{r})\right)^2$
converges absolutely. So, the $c_{ij}(\bs{r})$ can be seen as functions in 
$\ell^2(\ZZ^2)$ and, by the Riesz-Fischer Theorem \cite[Thm.\ 23.3]{Heuser}, 
each of the Fourier series 
$\sum_{\bs{r}\in\ZZ^2} c^{}_{ij}(\bs r) e^{-2\pi i \bs{k}\bs{r}}$
converges to a function in $L^2(\RR^2/\ZZ^2)$ in the $L^2$-norm. The limit
is independent of the order of summation, and convergence is actually
also pointwise, almost everywhere. H\"older's inequality gives
$L^2\left(\RR^2/\ZZ^2\right)\subset L^1\left(\RR^2/\ZZ^2\right)$, and
combining these periodic functions with the appropriate
phase shifts as implied by (\ref{domauto}) results in a 
function with lattice of periods 
$\Gamma=\{ (h,k)\in\ZZ^2\mid h+k \mbox{ even}\}$ which is
certainly in $L^1(\RR^2/\Gamma)$, so the 
Radon-Nikodym theorem \cite{RS} leads to the result stated. \qed \smallskip

{\sc Remark}: One may assign complex weights $h_1,h_2$ to the scatterers on 
the two dominoes without changing the spectral type. 
This results in the pure point part
\begin{equation}
   \left(\hat{\gamma}_{\omega}\right)_{pp} \; = \; 
         \frac{1}{4}\sum_{(h,k)\in\ZZ^2}\left| \rho_1 h_1+
                          (-1)^{h+k}\rho_2 h_2 \right|^2 \delta_{(h,k)} \, .
\end{equation}
This applies analogously to the next example.

\begin{figure}[ht]
\centerline{\epsfysize=6cm \epsfbox{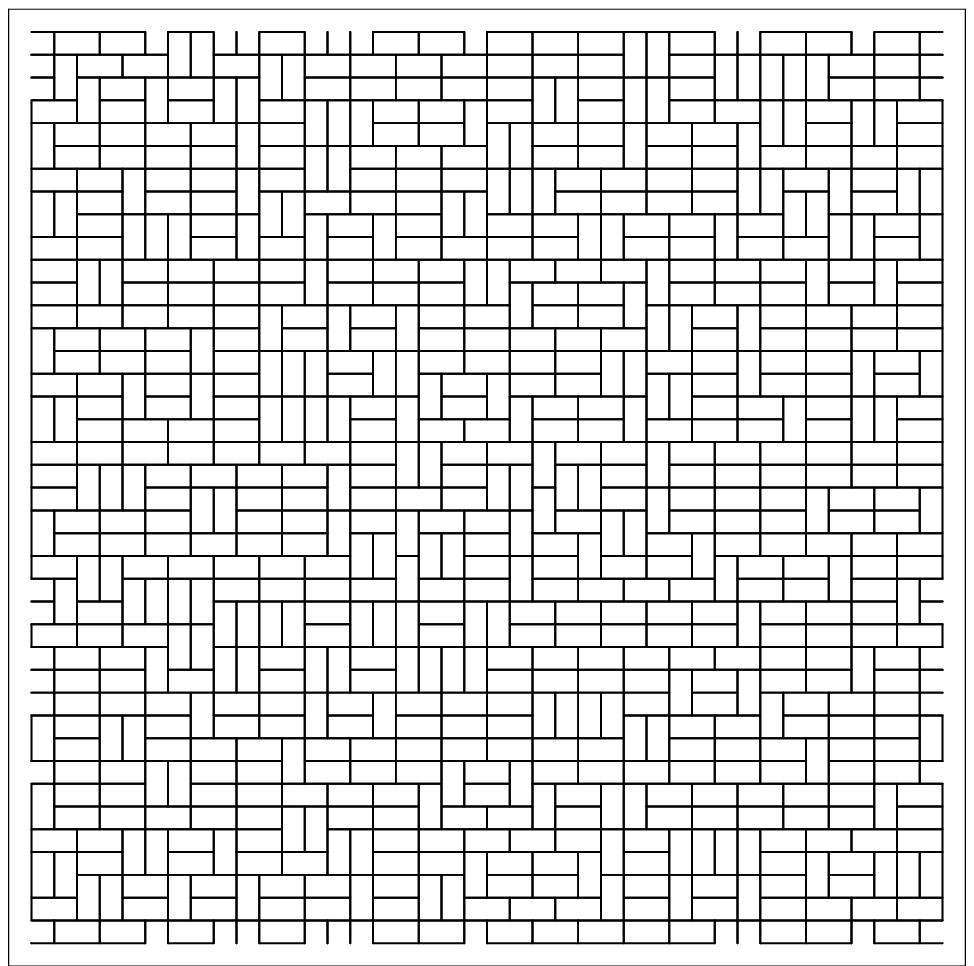} \hspace{1cm} 
\epsfysize=6cm \epsfbox{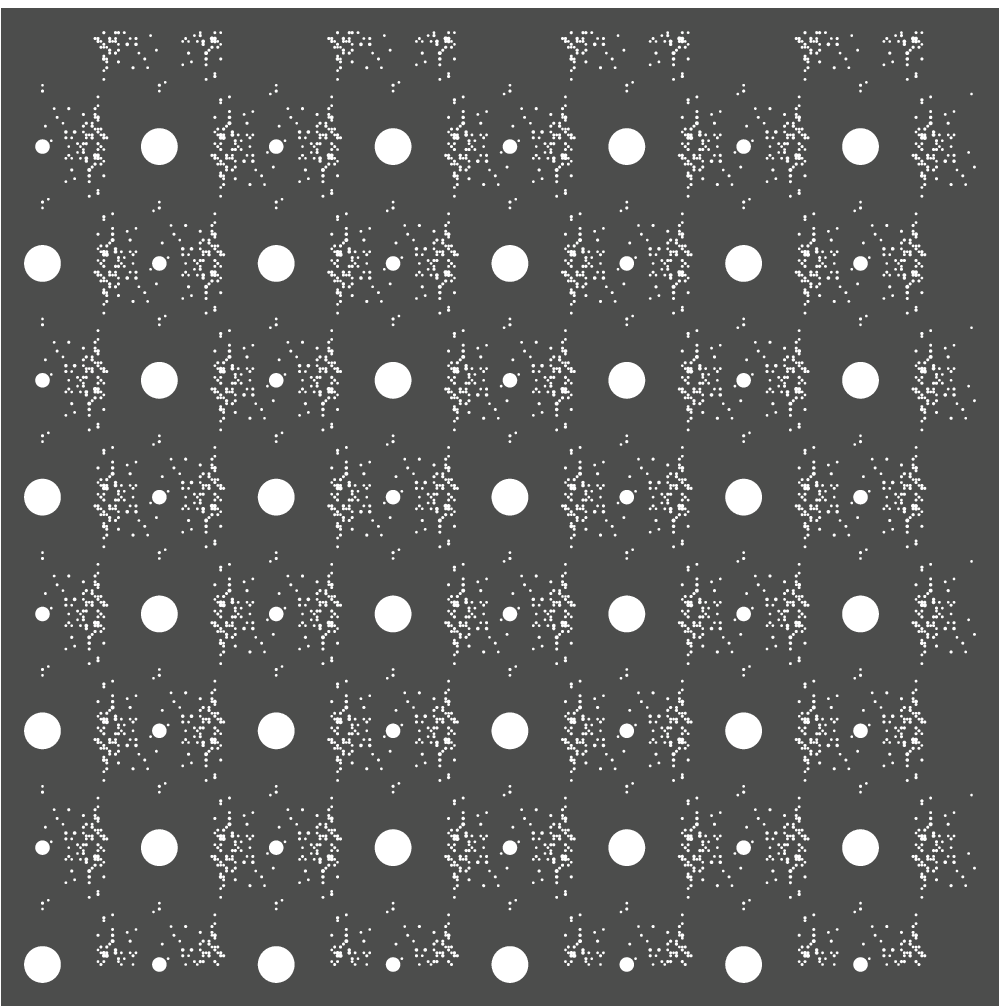}}
\caption{\label{domino} Typical tiling for $\rho_2^{}=0.3$ (left) and its 
diffraction image (right). The scatterers are located at the centre of the 
tiles.}
\end{figure}

One has to be aware that the pure point part (\ref{domauto}) does not display 
the correct (statistical) symmetry of the system. Away from the point of 
maximum entropy (which is $\rho_1^{}=\rho_2^{}=1/2$), 
the tiling is no longer fourfold symmetric, as still indicated
by the point part, but the twofold symmetry is only displayed in the diffuse 
background. This can be seen in Figure~\ref{domino}. The $pp$ part is calculated
from the exact expression and the Bragg peaks are represented by white circles 
with area proportional to the intensity. The $ac$ part was calculated 
numerically by means of standard FFT, because this is simpler than using the 
exact expression for the correlation functions.

In the special case of only one domino orientation remaining, the scatterers 
distribution is a Dirac comb on a rectangular lattice. Using Poisson's 
summation 
formula, the diffraction spectrum is a Dirac comb on the reciprocal rectangular 
lattice. In such a limit, the diffuse background accumulates at the extra 
positions and converges vaguely to a point measure that completes the square 
lattice arrangement to the proper rectangular one.

\subsubsection*{Lozenge tiling}

A lozenge is a rhombus with side 1, smaller angle $\pi/3$, and vertices in the 
triangular lattice $\Gamma=A_2^{}/\sqrt{2}$ with minimal distance $1$. This 
tiling can be mapped on a dimer configuration on the honeycomb packing. 
\begin{figure}[ht]
  \centerline{\epsfysize=4cm \epsfbox{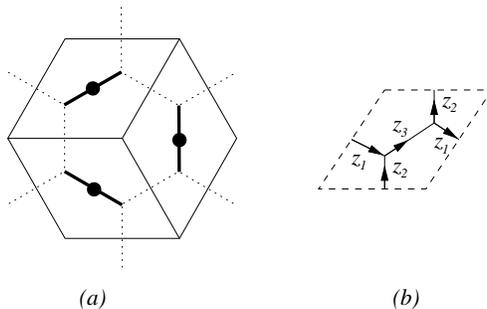}}
  \caption{\label{hexlattice}(a)~Lozenge tiling and the dimer configuration of 
  the honeycomb packing. (b)~Elementary cell for the weighted adjacency matrix.}
\end{figure}
The different tile densities are nonvanishing if 
$z_i^{}>|z_j^{}-z_k^{}|$, $i,j,k$ 
pairwise different \cite{Richard}; at equality, the system undergoes a phase 
transition of Kasteleyn type \cite{Kast2} with only one lozenge orientation 
remaining. This trivial case shall be excluded in the sequel. 
With Dirac unit measures on the tile centres, the support of the scatterers 
is given by a Kagom\'e grid of minimal vertex distance $1/2$. The adjacency 
matrix then has entries that are $2$ by $2$ matrices themselves, describing the 
elementary cells of the packing (see Figure \ref{hexlattice}). By wrapping the 
graph on a torus, we may transform it to block diagonal form with elements
\begin{equation}
  \bs{\lambda}(\varphi_1,\varphi_2) \; = \;
    \begin{pmatrix} 0 & 
       -(z_1^{}e^{-i\varphi_1}_{}+z_2^{}e^{-i\varphi_2}_{}+z_3^{}) \\
       z_1^{}e^{i\varphi_1}_{}+z_2^{}e^{i\varphi_2}_{}+z_3^{} 
       & 0 \end{pmatrix}\, .
\end{equation}
If we introduce a coordinate system with $\hat{x}=(1,0)^t_{}$ and 
$\hat{y}=1/2(1,\sqrt{3}\,)^t_{}$, we can use the above notation for the 
difference vectors of the elementary cells. Denoting the left and right site 
of the elementary cell by $L$ and $R$ we see that 
$[x,y]_{LL}^{}=[x,y]_{RR}^{}=0$. In the infinite size 
limit, the remaining matrix elements are given by
\begin{equation}
    [x,y|z_1^{},z_2^{},z_3^{}]_{LR} \; = \; 
         \frac{1}{4 \pi^2}\int_0^{2 \pi}\int_0^{2
         \pi}\frac{e^{i(\varphi_1^{}x+\varphi_2^{} y)}}
             {z_1^{}e^{-i \varphi_1^{}}+z_2^{}e^{-i \varphi_2^{}}+z_3^{}}
             d\varphi_1^{}d\varphi_2^{}
\end{equation}
(compare with \cite{Ken1}). Let $v=e^{-i \varphi_1^{}}$ and 
$w=e^{-i\varphi_2^{}}$. Then
\begin{equation} \label{lozcoup}
  [x,y|z_1^{},z_2^{},z_3^{}]_{LR} \; = \;
         \frac{1}{4 \pi^2}\int_{S^1\times S^1}\frac{v^{-x}w^{-y}}
               {z_1^{} v+z_2^{} w+z_3^{}}\frac{dv}{iv}\frac{dw}{iw} \, .
\end{equation}
As was already observed by Kenyon \cite{Ken1} for the isotropic case, the 
coupling
function has all the symmetries of the graph: interchanging $v$ and $w$ or the
substitution $(v,w)\to(w^{-1}_{},vw^{-1}_{})$ and combinations of these let the 
integral invariant, i.e.\
\begin{equation} \label{sym}
  \begin{array}{rcl}
       [x,y\mid z_1^{},z_2^{},z_3^{}]_{LR}
        & = & [x,-x-y-1\mid z_1^{},z_3^{},z_2^{}]_{LR} \\
  =\;  [y,x\mid z_2^{},z_1^{},z_3^{}]_{LR}
        & = & [y,-x-y-1\mid z^{}_2,z^{}_3,z^{}_1]_{LR} \\
  =\;  [-x-y-1,x\mid z_3^{},z_1^{},z_2^{}]_{LR}
        & = & [-x-y-1,y\mid z_3^{},z_2^{},z_1^{}]_{LR} \, .
  \end{array}
\end{equation}
We evaluate one integration in (\ref{lozcoup}) explicitly for $x\leq -1$. 
The other values can be obtained by (\ref{sym}). This is a direct 
generalization of \cite{Ken1} to the case of arbitrary activities. One gets
\begin{equation}
    [x,y|z_1^{},z_2^{},z_3^{}]_{LR} \; = \;
         \frac{i}{2 \pi} (-z_1^{})^x \int_{e^{i
           \varphi_0^{}}}^{e^{i(2\pi-\varphi_0^{})}} w^{-y-1}
           \left(z_2^{}+z_3^{}w\right)^{-x-1}dw,
\end{equation}
with $\varphi_0^{}=\arccos\frac{z_1^2-z_2^2-z_3^2}{2 z_2^{}z_3^{}}$. 
One easily finds the possible distance vectors of the scatterers. Away from 
the phase transition points, we have for the constant part of the 
autocorrelation measure for the scatterers (one per lozenge)
\begin{equation} \label{lozauto}
\begin{split}
   \left(\gamma_{\omega}^{}\right)_{const}^{} \; = \;
        \frac{2}{\sqrt{3}}
        \sum_{(x,y)\in \Gamma}\Bigl(&\left(\rho_1^2+\rho_2^2+\rho_3^2\right)
        \delta_{(x,y)}^{} \\ 
     &+2\rho_1^{}\rho_2^{}\delta_{\left(\frac{2x+1}{2},
        \frac{2y+1}{2}\right)}^{}+2\rho_1^{}\rho_3^{}
        \delta_{\left(\frac{2x+1}{2},y\right)}^{}+
         2\rho_2^{}\rho_3^{}\delta_{\left(x,\frac{2y+1}{2}\right)}^{}\Bigr).
\end{split}
\end{equation} 
The reciprocal lattice $\Gamma^*$ is spanned by the vectors 
$\left(1,-\frac{1}{\sqrt{3}}\right)^t$ 
and $\left(0,\frac{2}{\sqrt{3}}\right)^t$. Using (\ref{auto2d}) we can state
\begin{theorem} \label{thm5}
  Under the above assumptions, the diffraction spectrum of the lozenge tiling 
  exists 
  with probabilistic certainty and consists of a pure point and an absolutely 
  continuous part, i.e.\ $\hat{\gamma}^{}_{\omega} =
  (\hat{\gamma}^{}_{\omega})_{pp}  +  (\hat{\gamma}^{}_{\omega})_{ac}\,$, with
\begin{equation}
   \left(\hat{\gamma}^{}_{\omega}\right)_{pp} \; = \; 
   \frac{4}{3}\sum_{(h,k)\in \Gamma^*}
   \left((-1)^h\rho_1^{}+(-1)^k\rho_2^{}+\rho_3^{}\right)^2 \delta_{(h,k)}^{}.
\end{equation}
  There is no singular continuous part, and $\hat{\gamma}^{}_{\omega}$
  is periodic with lattice $2\Gamma^*$.
\end{theorem}
{\sc Proof}: The pure point part is simply the Fourier transform of 
(\ref{lozauto}), again calculated by means of Poisson's summation formula. 

\begin{figure}[ht]
\centerline{\epsfysize=6cm \epsfbox{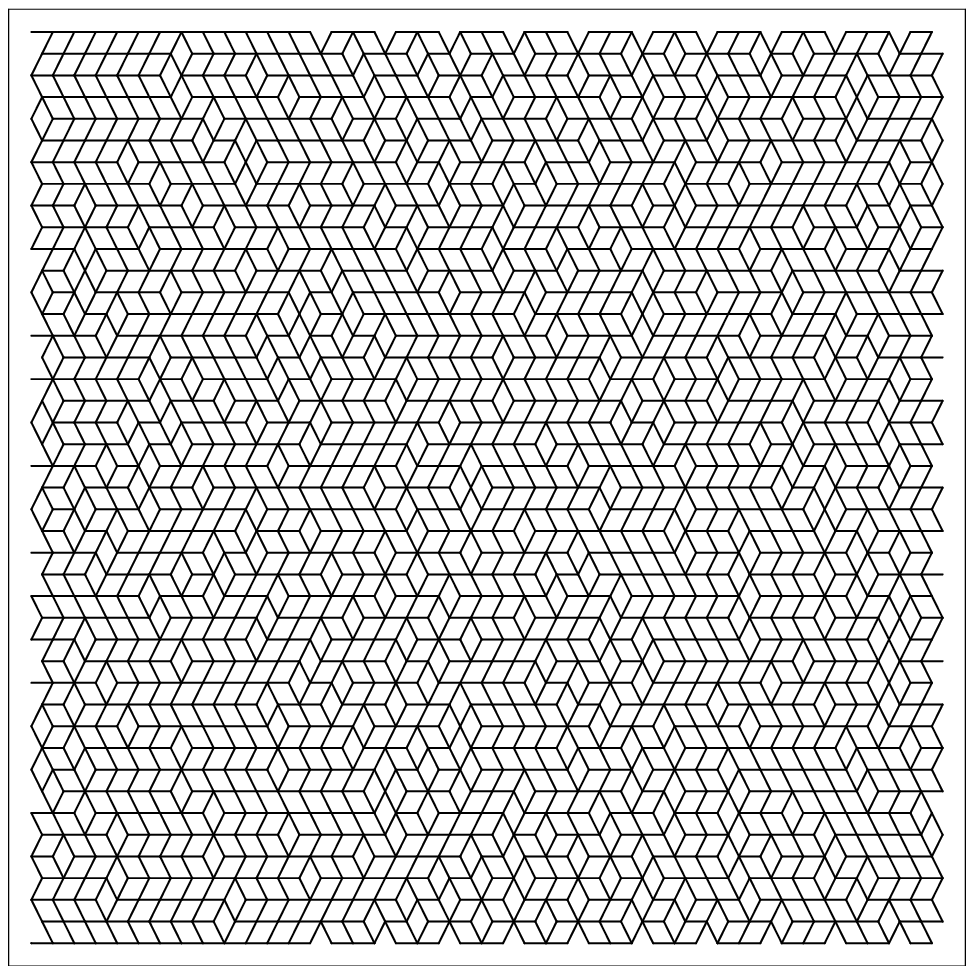} \hspace{1cm} 
\epsfysize=6cm \epsfbox{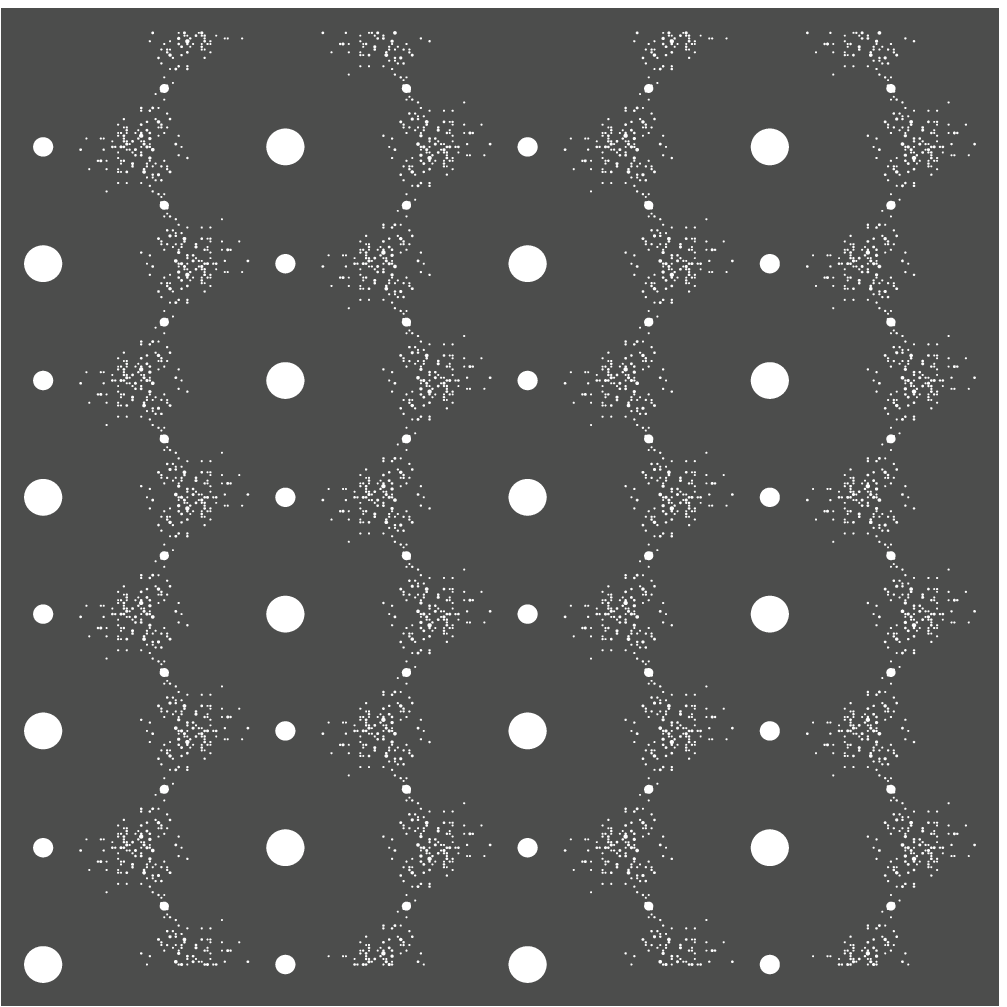}}
\caption{\label{hex} Typical tiling for $\rho_2^{}=0.24$ and 
$\rho_1^{}=\rho_3^{}=0.38$ (left) and its diffraction image (right). 
The scatterers are located at the centres of the tiles.}
\end{figure}

As before, we will now show that the remaining part of 
$\hat{\gamma}^{}_{\omega}$ 
converges to a periodic $L^1$-function and thus represents an absolutely 
continuous measure. Let us start with the case $x\leq -1$ (and $y$ arbitrary, 
but fixed) and show that 
$[x,y|z_1^{},z_2^{},z_3^{}]_{LR} = \mathcal{O}\left({|x|^{-1}}\right)$
as $x\to - \infty$. The necessary extension to the complete asymptotic 
behaviour will later follow from Eq.~(\ref{sym}). For now, and for fixed values
of the activities in the admitted range, we have
\begin{align}
   I \; = \; \left|\,[x,y|z_1^{},z_2^{},z_3^{}]_{LR}\,\right| &\;\leq\;
   \frac{z_1^x}{2\pi}
   \int_{e^{i\varphi_0^{}}}^{e^{i(2\pi-\varphi_0^{})}}|z_2^{}+z_3^{}
        w|^{-x-1}|dw| \nonumber\\
       &\;=\;
  \frac{z_1^x}{\pi}\int_{\varphi_0^{}}^{\pi}\left(z_2^2+z_3^2+2z_2^{}z_3^{}
         \cos\vartheta\right)^{\frac{-x-1}{2}} d\vartheta.
\end{align}
We first bound $f(\vartheta)=(z_2^2+z_3^2+2 z_2^{} z_3^{}\cos\vartheta)/z_1^2$ 
by a straight line $\tilde{f}(\vartheta)$, i.e.\ we look for 
$f(\vartheta)\leq\tilde{f}(\vartheta)$ on the interval $[\varphi^{}_0,\pi]$.
Note that $f(\vartheta)$ can only have
one flex point at $\vartheta=\pi/2$ in $[0,\pi]$. One has to distinguish two 
cases:
\begin{enumerate}
\item $\cos\varphi_0^{}\geq(\pi-\varphi_0^{})\sin\varphi_0^{} - 1$: 
         Choose $\tilde{f}_1(\vartheta)
          =(\vartheta-\varphi_0^{})f'(\varphi_0^{})+f(\varphi_0^{})$. 
        The condition on $\varphi_0^{}$ implies
        $\tilde{f}_1(\pi)\geq f(\pi)$. 
        Differentiating $f(\vartheta)-\tilde{f}_1(\vartheta)$
        with respect to $\vartheta$ yields 
        $f'(\vartheta)-\tilde{f}_1'(\vartheta) 
        = f'(\vartheta) - f'(\varphi^{}_0)$ and this vanishes if
        $\vartheta = \varphi^{}_0$ (maximum) or if 
        $\vartheta = \pi-\varphi_0^{}$ 
        (minimum). Thus $f(\vartheta)\leq\tilde{f}_1(\vartheta)$.
\item $\cos\varphi_0^{}<(\pi-\varphi_0^{})\sin\varphi_0^{} - 1$: 
        Choose $\tilde{f}_2(\vartheta)
        =\frac{f(\pi) - 1}{\pi-\varphi_0^{}}\vartheta+ 
        \frac{\pi-\varphi_0^{}f(\pi)}
        {\pi-\varphi_0^{}}$ (the line connecting $f(\varphi_0^{})=1$ and 
        $f(\pi)$). 
        Obviously $\tilde{f}'_2(\pi)\leq f'(\pi)$. Because of the angle 
        condition, 
        we further have $f'(\varphi_0^{})\leq \tilde{f}'_2(\varphi_0^{})$. 
        As there is only one flex point,
        we conclude that $f(\vartheta)\leq\tilde{f}_2(\vartheta)$.
\end{enumerate}
Choose $f(\vartheta)$ as $f_1(\vartheta)$ or $f_2(\vartheta)$ according to the
previous distinction. If $c=|\tilde{f}'(\vartheta)|$, we get
$0<c\leq\frac{1}{\pi-\varphi^{}_0}\left(1-\frac{(z^{}_2-z^{}_3)^2}{z_1^2}\right)
<\frac{1}{\pi-\varphi^{}_0}$. Consequently, we can estimate
\begin{align}
   \pi z^{}_1 I&\;\leq\;\int_{\varphi_0^{}}^{\pi}
   \left(1-c(\vartheta-\varphi_0^{})\right)^{\frac{-x-1}{2}}
   d\vartheta \nonumber \\
   &\;<\;\frac{1}{c}\int_0^1 (1-t)_{}^{\frac{-x-1}{2}} \, dt  \\
   &\;=\; \frac{2}{c(1-x)}
    \;=\;\mathcal{O}\left(\frac{1}{|x|}\right) . \nonumber
\end{align}

Now, we can use the symmetry relations of (\ref{sym}) and obtain the asymptotic
behaviour
$[x,y|z_1^{},z_2^{},z_3^{}]_{LR}=\mathcal{O}\left({(|x|+|y|)^{-1}}\right)$, 
whenever $|\bs r| \to\infty$, compare \cite{Ken1}. 
The rest of the argument is very similar to the domino case and need
not be repeated here, the periodicity statement follows again
constructively.   \qed \smallskip

Let us remark that an analogous scenario, with the same type
of result, occurs for the more complicated dart-rhombus
random tiling, see \cite{Hoeffe} for details.

\subsection*{Addendum: The two-dimensional Ising model}

For the sake of completeness, we add an application of the probably best 
analyzed model in statistical physics, the 2D Ising model without external 
field. It may be regarded as a lattice gas on $\ZZ^2$, compare \cite{Simon}, 
with scatterers of strength $s^{}_{(i,j)}\in\{1,0\}$. The partition function 
in the spin-formulation ($\sigma^{}_{(i,j)}\in\{+1,-1\}$) reads as follows
\begin{equation}
   \mathcal{Z} \; = \; \sum_{\{\bs{\sigma}\}}\exp\left(\sum_{(i,j)}
   \sigma^{}_{(i,j)}\left(K^{}_1\sigma^{}_{(i+1,j)}+
                       K^{}_2\sigma^{}_{(i,j+1)}\right)\right) \, ,
\end{equation}
where we sum over all configurations $\{\bs{\sigma}\}$. We consider the 
ferromagnetic case with coupling constants $K_i=J_i/(k_B T)>0$,  temperature 
$T$ and Boltzmann's constant $k_B$. The model undergoes a phase transition at 
$k:=\left(\sinh(2 K_1)\sinh(2 K_2)\right)^{-1}=1$. It is common knowledge that 
in the regime with coupling constants smaller than the critical ones 
(corresponding to $T>T_c$) the ergodic equilibrium state with vanishing 
magnetization $m$ is unique, whereas above ($T<T_c$) there exist two extremal 
equilibrium states, which are thus ergodic \cite[Ch.~III.5]{Simon}. In this 
case, we assume to be in the extremal state with positive magnetization 
$m=(1-k^2)^{1/8}$.

The diffraction properties of the Ising model can be extracted from the known 
asymptotic behaviour \cite{McCoy,Wu} of the autocorrelation coefficients. We
first state the result for the isotropic case ($K_1=K_2=K$) and
comment on the general case later.
\begin{prop} \label{ising}
  Away from the critical point, the diffraction spectrum of the Ising lattice 
  gas almost surely exists, is $\ZZ^2$-periodic and consists of 
  a pure point and an absolutely continuous part with continuous density.
  The pure point part reads
\begin{enumerate}
\item $T>T_c$: $(\hat{\gamma}_{\omega})_{pp}=\frac{1}{4}
               \sum_{\bs{k}\in\ZZ^2}\delta_{\bs{k}}^{}$
\item $T<T_c$: $(\hat{\gamma}_{\omega})_{pp}=\rho^2
               \sum_{\bs{k}\in\ZZ^2}\delta_{\bs{k}}^{}$,
\end{enumerate}
  where the density $\rho$ is the ensemble average of the number of scatterers 
  per unit volume.
\end{prop}
{\sc Proof}: First, note that $s^{}_{(i,j)}=(\sigma^{}_{(i,j)}+1)/2$ and 
thus $\langle\sigma^{}_{(i,j)}\rangle = m=2\rho-1$, so $\rho$ varies between
$1$ and $1/2$. The asymptotic correlation function of two spins at distance 
$R=\sqrt{x^2+y^2}\,$ (as $R\to\infty$) is \cite{McCoy}
\begin{equation} \label{asymp}
    \langle \sigma^{}_{(0,0)}\sigma^{}_{(x,y)}\rangle \; \simeq \;
\begin{cases}
        c_1\frac{e^{-R/c_2}}{\sqrt{R}},       & T>T_c \\
        m^2+c_3\frac{e^{-2R/c_2}}{R^2},\quad  & T<T_c,
\end{cases}
\end{equation}
with constants $c_1,c_2$ and $c_3$ depending only on $K$ and
$T$, see also \cite[p.\ 51]{KS} and references given there
for a summary.
The pure point part $(\hat{\gamma}_{\omega}^{})_{pp}$ results directly 
from the Fourier transform of the constant part of $\gamma_{\omega}^{}$ as 
derived from the asymptotics of $\langle s_{(0,0)}^{} s_{(x,y)}^{} \rangle =
(\langle\sigma_{(0,0)}^{}\sigma_{(x,y)}^{}\rangle + 2 m + 1)/4$.

Here, already $\sum_{(x,y)\in\ZZ^2} e^{-R/c_2}/\sqrt{R}$ and 
$\sum_{(x,y)\in\ZZ^2} e^{-2R/c_2}/R^2$ converge absolutely, so we can view
the corresponding correlation coefficients as functions in $L^1(\ZZ^2)$.
Their Fourier transforms (which are uniformly converging Fourier series) 
are continuous functions on $\RR^2/\ZZ^2$, see \cite[\S 1.2.3]{Rudin1},
which are then also in $L^1(\RR^2/\ZZ^2)$.
Applying the Radon-Nikodym theorem finishes the proof. \qed \smallskip

{\sc Remark}: At the critical point, the correlation function
$\langle \sigma_{(0,0)}^{}\sigma_{(x,y)}^{}\rangle$ is asymptotically 
proportional to $R^{-1/4}$ as $R\to\infty$ \cite{Wu,KS}. Again, taking out
first the constant part of $\gamma_{\omega}^{}$, we get the same pure
point part as in Prop.\ \ref{ising} for $T>T_c$. However, for the
remaining part of $\gamma_{\omega}^{}$, both our previous arguments fail. 
Nevertheless, using a theorem of Hardy \cite[p.~97]{Bromwich}, 
we can show that the corresponding Fourier series still converges
for $\bs{k}\not\in \ZZ^2$ (a natural order of summation is given
by shells of increasing radius).

{}For $\bs{k}\in\ZZ^2$, where the Bragg peaks reside, the series diverges.
But this can neither result in further contributions to the Bragg peaks
(the constant part of $\gamma_{\omega}^{}$ had already been taken care of)
nor in singular continuous contributions (because the points of divergence
form a uniformly discrete set). So, even though the series diverges for 
$\bs{k}\in\ZZ^2$, it still represents (we know that 
$\hat{\gamma}^{}_{\omega}$ exists) a function in $L^1(\RR^2/\ZZ^2)$ and 
hence the Radon-Nikodym density of an absolutely continuous background.
On the diffraction image, we thus can see, for any temperature, Bragg peaks 
on the square lattice and a $\ZZ^2$-periodic, absolutely continuous background 
concentrated around the peaks (the interaction is attractive). 
At the critical point, the intensity of the diffuse scattering diverges when 
approaching the lattice positions of the Bragg peaks. 

The same arguments hold in the anisotropic case, where the asymptotics 
still conforms to Eq.~(\ref{asymp}) and the above, if $R=R(x,y)$ is replaced by
the formula given in \cite[Eq.\ 2.6]{Wu}. The pure point part is again
that of Prop.~\ref{ising} with fourfold symmetry, while (as in the case of the 
domino tiling) the continuous background breaks this symmetry if $K_1\neq K_2$.
Let us finally remark that a different choice of the scattering strengths 
(i.e.\ $\pm 1$ rather than $1$ and $0$) would result in the extinction 
of the Bragg peaks in the disordered phase ($T>T_c$), but no choice does so in 
the ordered phase ($T<T_c$).

\subsection*{Outlook}

The diffraction of crystallographic and of perfect quasi-crystallographic
structures is well understood. The main aim of this article was to begin to 
counterbalance this into the direction of certain stochastic arrangements of
scatterers, and to random tiling arrangements in particular. This
requires a careful investigation of the diffuse background and clear concepts 
about absolutely versus singular continuous contributions. Such problems are 
naturally studied via spectral properties of unbounded complex measures which 
we did for a number of simple, but relevant examples.

While stochastic cuboid tilings would generically display a singular continuous
contribution in the diffraction image, our results on the domino and the 
lozenge tilings show that no such contributions exist there. However, we do 
not think that this is a robust result. In fact, the natural next step would 
be an extension to planar random tilings with quasi-crystallographic symmetries
such as 8-, 10- or 12-fold. Then, according to the folklore results, one should 
expect a purely continuous diffraction spectrum (except for the trivial Bragg 
peak at $\bs{k}=0$ which merely reflects the existing natural density of the 
scatterers). This spectrum should then split into an $ac$ and an $sc$ part.

But is the replacement of Bragg peaks by $sc$ peaks significant? Scaling 
arguments \cite{Henley} and numerical calculations \cite{Dieter} indicate that 
the exponents of important $sc$ peaks can be extremely close to $1$ which 
means that their distinction from Bragg peaks is almost impossible in practice. 
Is it possible that this is one reason why structure refinement is so difficult
for real decagonal quasicrystals? This certainly demands further thought, but 
it is not clear at the moment to what extent a rigorous treatment is possible.

Finally, except for Bernoulli type systems \cite{BM1} and for extensions by 
means of products of measures, we have not touched the ``real'' diffraction 
issues in 3-space. 
One reason is that we presently do not know of any interesting model that can be
solved exactly (let alone rigorously), another is that, for random tiling models
relevant to real quasicrystals, one expects bounded fluctuations (again, due to
heuristic scaling arguments, see \cite{Henley} and references therein).
Consequently, the diffraction spectra should typically show Bragg peaks plus
an absolutely continuous background. We hope to report on some progress soon.

\subsection*{Acknowledgements}

It is our pleasure to thank Joachim Hermisson, Anders Martin-L\"of, 
Robert V.\ Moody, Wolfram Prandl and Martin Schlott\-mann 
for several helpful discussions and comments. We are grateful to 
Aernout C.\ D.\ van Enter for helpful advice and for setting us right in the 
Addendum. This work was
supported by the Cusanuswerk and by the German Research Council (DFG).

\end{document}